\documentclass[twocolumn,prb,amsmath,amssymb]{revtex4}

\usepackage{graphics}
\usepackage{graphicx}
\usepackage{dcolumn}
\usepackage{bm}

\begin{document}

\title{Variable-range hopping in quasi-one-dimensional electron
crystals}

\author{M.~M.~Fogler}

\affiliation{Department of Physics, University of
California San Diego, La Jolla, California 92093}

\author{S.~Teber and B.~I.~Shklovskii}

\affiliation{William I. Fine Theoretical Physics Institute, University of
Minnesota, Minneapolis, Minnesota 55455}

\date{\today}

\begin{abstract}

We study the effect of impurities on the ground state and the
low-temperature Ohmic dc transport in a one-dimensional chain and
quasi-one-dimensional systems of many parallel chains. We assume that
strong interactions impose a short-range periodicicity of the electron
positions. The long-range order of such an electron crystal (or
equivalently, a $4 k_F$ charge-density wave) is destroyed by impurities,
which act as strong pinning centers. We show that a three-dimensional array
of chains behaves differently at large and at
small impurity concentrations $N$. At large $N$, impurities divide the
chains into metallic rods. Additions or removal of electrons from such
rods correspond to charge excitations whose density of states exhibits
a quadratic Coulomb gap. At low temperatures the conductivity
is due to the variable-range hopping of electrons between the rods. It obeys
the Efros-Shklovskii (ES) law $-\ln \sigma \sim (T_{\rm ES} / T)^{1/2}$.
$T_{\rm ES}$ decreases as $N$ decreases, which leads to an exponential
growth of $\sigma$. When $N$ is small, the metallic-rod (also
known as ``interrupted-strand'') picture of the ground state survives
only in the form of rare clusters of atypically short rods. They are the
source of low-energy charge excitations. In the bulk of the crystal the
charge excitations are gapped and the electron crystal is pinned
collectively. A strongly anisotropic screening of the Coulomb
potential produces an unconventional linear in energy Coulomb
gap and a new law of the variable-range hopping conductivity $-\ln\sigma
\sim (T_1 / T)^{2/5}$. The parameter $T_1$ remains constant over a
finite range of impurity concentrations. At smaller $N$ the $2/5$-law is
replaced by the Mott law, $-\ln \sigma \sim (T_{M} / T)^{1/4}$. In the
Mott regime the conductivity gets suppressed as $N$ goes down. Thus, the
overall dependence of $\sigma$ on $N$ is nonmonotonic. In the case of a
single chain, the metallic-rod picture applies at all $N$. The
low-temperature conductivity obeys the ES law, with log-corrections, and
decreases exponentially with $N$. Our theory provides a qualitative
explanation for the transport properties of organic charge-density wave
compounds of TCNQ family.

\end{abstract}

\maketitle

\section{Introduction}
\label{Sec:Introduction}

In recent years electron transport in quasi-one-dimensional (quasi-1D)
systems moved into focus of both fundamental and applied research.
Quantum wires, nanotube ropes, conducting molecules, {\it etc.\/},
are being examined as possible elements of miniature electronics
devices. In parallel, discovery of quasi-1D structures termed
``stripes'' in correlated electron systems (high-$T_c$ cuprates, quantum
Hall devices, {\it etc\/}), invigorates efforts to understand
unconventional phases in two and three dimensions starting from models
of weakly coupled 1D chains.

Experimentally, the low-temperature conductivity $\sigma(T)$ of quasi-1D
systems is often of the insulating type. Its temperature dependence
gives information about the nature of charge excitations. For example,
the activated dependence $-\ln\sigma(T) \propto 1 / T$ indicates a gap in
the spectrum. In quasi-1D systems such a gap commonly arises from the
Mott-Peierls mechanism,~\cite{YuLu,Heeger_84} where the commensurability
with the host lattice is crucial. Yet there are many 1D and quasi-1D
systems where commensurability plays a negligible role. In this
situation the jellium model (an electron gas on a positive compensating
background) is a good approximation. This is the kind of systems we
study in this paper. We will show that their low-temperature transport
is dominated by a variable-range hopping (VRH), which leads to a slower
than exponential $T$-dependence~\cite{Mott_book,ES84} of the
conductivity. Our theory may apply to a number of systems, both
naturally occuring and man-made. Prototypical 1D examples are individual
quantum wires an carbon nanotubes. Stripe
phases,~\cite{Fogler_review,Stripes_in_oxides} quantum wire arrays in
heterojunctions,~\cite{Mani_92} carbon nanotube films,~\cite{deHeer_95}
and atomic wires on silicon surface,~\cite{Himpsel} are two-dimensional
(2D) examples. In three dimensions (3D), an important and well studied
class of quasi-1D compounds are charge-density
waves,~\cite{Gruner,Organics_review,ECRYS} (CDW).

\begin{figure}
\includegraphics[width=3.4in]{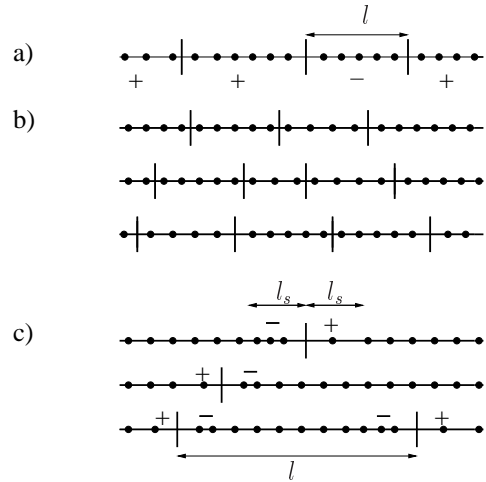}
\vspace{0.1in}
\caption{Pinned 1D and quasi-1D systems on a uniformly charged positive
background. Dots and tick marks label the positions of electrons and
impurities, respectively. (a) 1D crystal. The $+$ and $-$ signs denote
the charge of the metallic rods, which have an average length of $l$.
(b) An array of decoupled chains in the case $l < l_s$. (c) An array of
coupled chains for the case $l > l_s$. The preferred arrangement of the
electrons on neigborng chains may depend on the exact geometry of the
system. In order not to complicate the drawing, we adopted the
convention where the ground state corresponds to the same horizontal
positions of electrons on each chain. The interchain interactions try to
diminish the deviations from this ground state leading to dipolar
distortions of a characteristic length $l_s$ around impurities.
\label{Pinned_CDW}}
\end{figure}

To characterize the strength of Coulomb correlations in a quasi-1D
system we define the dimensionless parameter $r_s \equiv a / 2 a_B$,
where $a$ is the average distance between electrons along the chain
direction and $a_B = \hbar^2 \kappa / m e^2$ is the effective Bohr
radius. The latter is expressed in terms of the dielectric constant of
the medium $\kappa$ and the electron band mass $m$. In practically all
known realizations of 1D and quasi-1D systems, $r_s$ exceeds unity,
often by orders of magnitude. Below we assume that $r_s \gg 1$. Under
this condition the dynamics of electrons can be treated semiclassically.
Neglecting quantum fluctuations altogether for a moment, we arrive at
the picture of electrons forming a classical 1D Wigner crystal in the
case of a single chain (Fig.~\ref{Pinned_CDW}a) or an array of such
crystals in 2D and 3D (Fig.~\ref{Pinned_CDW}b and c), with the period
along the chain equal to $a$. Formation of the crystal enables the
electrons to minimize the energy of their mutual Coulomb repulsion. In
order to correctly assess the role of quantum fluctuations (zero-point
motion) in such crystals, one has to take into account two
circumstances. First is the unavoidable presence of random impurities
that act as pinning centers. Second is a finite interchain coupling (in
2D and 3D systems). Either one is sufficient to make the quantum
fluctuations of electron positions bounded. For example, in the case of
a single chain, the slow growth of the zero-point motion amplitude with
distance~\cite{Glazman_92,Schulz_93} is terminated at the nearest strong
pinning center. In higher dimensions, the zero-point motion amplitude is
finite even without impurities because of the interchain interaction. In
fact, renormalization group
approaches~\cite{Interchain_coupling_studies} indicate that the Wigner
crystal or, equivalently, $4 k_F$-CDW is the true ground state of a
system of weakly coupled chains starting already from rather modest
$r_s$. In all situations, the net effect of quantum fluctuations is to
slightly renormalize the the bare impurity strength and/or bare
interchain coupling. The calculation of renormalized parameters is
possible via the standard bosonization
technique.~\cite{Gogolin_book} For the treatment of, e.g., 1D case,
one can consult
Refs.~\onlinecite{Giamarchi_88,Glazman_92,Kane_92,Schulz_93}. Below such
a renormalization is assumed to be taken into account and it is not
discussed further. Henceforth we will often refer to the systems we
study as {\it electron crystals\/}.

Due to impurity pinning, at zero temperature and in the limit of small
electric field (Ohmic regime) the conductivity of the electron crystal
vanishes. This behavior is common for all pinned systems. It motivated a
large body of work~\cite{Blatter_94,ECRYS} devoted to mechanisms of {\it
nonlinear\/} transport that become possible in the presence of a finite
electric field, in particular, creep and sliding. In the present
context, such mechanisms would involve a collective motion of large
numbers of electrons that overcome pinning barriers either by thermal
activation or by quantum tunneling. It has been understood, however, on
some qualitative level,~\cite{Blatter_94} that if compact charge
excitations are allowed by the topology of the system, then such
excitations would dominate the response at low temperatures and would
give rise to a nonzero {\it Ohmic\/} conductivity at $T > 0$. Below we
will demonstrate that this is indeed the case in the electron crystals.
We clarify the nature of the compact low-energy charge excitations and
propose a theory of their low-temperature Ohmic transport that
consistently addresses the role of long-range Coulomb
interactions.~\cite{Comment_on_noninteracting}

In this paper we are focused exclusively on the charge transport and
ignore any effects related to the spin degree of freedom. This is
legitimate for $r_s \gg 1$ because electron are tightly localized at the
sites of the classical Wigner crystal and the energy of their
spin-dependent exchange interaction is exponentially small.

We will assume that impurities that pin the crystal are strong enough to
enforce preferred order of electrons nearby or, in the CDW terminology,
the preferred phase. The relation between the phase $\phi$ and the
elastic displacement of the crystal $u$ is $\phi = -(2 \pi / a) u$. In
Fig.~\ref{Pinned_CDW} impurities are shown by vertical tick marks and it
is assumed that they interact with nearby electrons by a repulsive
potential comparable in magnitude to the Coulomb interaction energy
$e^2/ \kappa a$ between nearest electrons on the chain. This condition
is sufficient to ensure that the impurity to act as a strong pinning
center. An example of such an impurity is an acceptor residing
on the chain. In the ground state one electron is bound to the acceptor
and the electron-acceptor complex (of total charge $e$) is built into
the crystal, i.e., it is positioned squarely in between the two closest
other electrons. One can say that the crystal contains a plastic
deformation --- a vacancy bound to the negatively-charged acceptor.
Overall, the region around the impurity is electrically neutral.

In the case of a single chain (Fig.~\ref{Pinned_CDW}a), strong
impurities divide the crystal into segments, which behave as individual
metallic rods. A charge can easily spread over the length of each rod,
while it has to tunnel through an impurity to move to a neighboring rod.
Each rod contains an integer number of electrons but the charge of the
positive background is random because of the assumed incommensurability.
In the ground state the distribution of the rods' total charges,
electrons plus background, is uniform between $-e/2$ and $+e/2$ ($\phi$
is between $-\pi$ and $\pi$). Larger charges cost more Coulomb energy
and correspond to charge excitations above the ground state. Transitions
between ground and excited states occur by discrete changes in the
number of electrons on the rods.

Let $\varepsilon$ be the minimal by absolute value change in the self-energy
of a given rod due to change of its charge by one unit. By
the self-energy we mean the Coulomb energy of interaction among the
electrons on the given rod, charges on all other rods held fixed. We
adopt the convention that the chemical potential (Fermi energy)
corresponds to the zero energy. In this case, $\varepsilon$ is
nonnegative (nonpositive) for addition (subtraction) of the electron. We
denote by $g_B(\varepsilon)$ the distribution function of $\varepsilon$
averaged over impurity positions. We refer to $g_B$ as the bare density
of states of charge excitations. In Sec.~\ref{Sec:1D} we show that
random distribution of charges of the rods creates a finite $g_B(0)$.
Small $\varepsilon$ come from rods with net charges close to $\pm e /
2$. These rods make possible a VRH conductivity at low temperatures.

Consider now a 3D system of parallel chains. Impurities with
concentration $N$ divide the chains into segments of average length $l =
1 / N a_\perp^{2}$, where $a_\perp^{2}$ is the area per chain in $y$-$z$ plane (we
assume the chains to be along the $x$-direction). We get two cases
distinguished by the relative importance of interchain interactions. In
the first case (Fig.~\ref{Pinned_CDW}b), the chains are far enough from
each other and/or the impurity concentration is large enough so that the
interchain coupling over the length $\sim l$ of a typical segment can be
neglected in comparison with its longitudinal compression energy. As a
result, the phases of different segments are uncorrelated and the system
again behaves as a collection of metallic rods. Polarizability of the
rods generates a strongly anisotropic dielectric constant. Like for a
single chain, in the ground state of the system, a finite bare density
of states $g_B$ at zero energy originates from random background charges
of the rods. Again this leads to the VRH at low temperatures.

In the other case (Fig.~\ref{Pinned_CDW}c), the concentration of
impurities is small and chains are strongly interlocked. The elastic
distortions are concentrated in small regions around individual
impurities (see below). Away from impurities the crystal possess a good
3D order. The true long-range order is however absent because of the
cumulative effect of small elastic distortions in a large volume. The
elastic displacement field $\bar{u}({\bf r})$ of the electron crystal
lattice away from impurities gradually varies in space. The length scale
where its variation is of the order of $a$ (variation in $\phi$ is of
the order of unity) is referred to as Larkin length. The Larkin length
is exponentially large,~\cite{Bergman_77,Efetov_77} effectively
infinite, because the Coulomb interactions make the crystal very rigid.
This will be discussed in more detail in Sec.~\ref{Sec:3D_clean}.

The region near a typical impurity has the following
structure.~\cite{Larkin_94} On one side of the impurity the chain is
compressed, which creates an excess negative charge; on the opposite
side, it is stretched resulting in a positive charge of the same
absolute value $\leq e/2$. The net charge of such a dipole is zero
(together with the vacancy of charge $-e$ and the acceptor of charge
$e$). The characteristic length of the distorted region is of the same
order as the length $l_s$ of the nonlinear topological
excitation~\cite{RBKT,BK,Comment_on_vacancy} of the pure system, the
$2\pi$-soliton. This is because the magnitude of local distortions in
$\phi$ are typically comparable in the two cases. The formal
definition of $l_s$ is~\cite{Brazovskii_91}
\begin{equation}
                    l_s = a_\perp / \sqrt{\alpha},
\label{l_s_def}
\end{equation}
where $\alpha = Y_\perp / Y_x \ll 1$ is the dimensionless anisotropy
parameter and $Y_x$, $Y_\perp$ are the longitudinal and transverse
elastic moduli of the electron crystal. The elasticity theory of the
crystal will be discussed in more detail in Sec.~\ref{Sec:3D_clean}.
Here we just mention that $\alpha$ varies from material to material,
e.g., $\alpha \sim 10^{-4}$ in KCP and in CDW organics, $\alpha \sim
10^{-2}$ in blue bronze.~\cite{Gruner,Organics_review}

At low temperatures all solitons are bound to acceptors with large
binding energies comparable to the creation energy of a free soliton $W
\sim e^2/ \kappa l_s$. In other words, there is a large energy gap for
creating charge excitations. Nevertheless, as shown in
Sec.~\ref{Sec:3D_clean}, finite $g_B$ at zero energy does exist in the
case $l \gg l_s$ as well. It comes from rare clusters of several closely
spaced impurities. Such clusters can be viewed as microscopic inclusions
of the $l \ll l_s$ phase (where $g_B$ is large).

In all cases outlined above, $g_B$ is not yet the actual density of
states of charge excitations. This is because the long-range Coulomb
interaction of charges at distant sites is not included in the
definition of $g_B$. We denote the true density of states of charge
excitations by $g(\varepsilon)$. Based on previous studies of other
insulating systems, such as doped semiconductors,~\cite{ES75,ES84} we
expect that long-range interactions generate a Coulomb gap in
$g(\varepsilon)$. This gap is soft, in the sense that $g(\varepsilon)$
vanishes only at the Fermi level $\varepsilon = 0$. Away from the Fermi
level, $g(\varepsilon)$ increases in a power-law fashion until it
saturates at the bare value $g_B$ at large enough $\varepsilon$
(cf.~Ref.~\onlinecite{ES84} and Fig.~10.4 therein). Note that $g$ is
different from the thermodynamical density of states. The latter does
not vanish despite Coulomb correlations, see Refs.~\onlinecite{ES84}
and~\onlinecite{ES85}.

In macroscopically {\it isotropic\/} electron systems the functional
form of the Coulomb gap depends on the number of dimensions. The density
of states behaves as $\varepsilon^2$ in 3D and as $|\varepsilon|$ in 2D.
In all dimensions, however, this leads to the the Efros-Shklovskii (ES)
law for the VRH conductivity in isotropic doped
semiconductors~\cite{ES75,ES84}
\begin{equation}
\sigma = \sigma_0 \exp[-(T_{\rm ES}/T)^{1/2}],
\label{DC}
\end{equation}
where $\sigma_0$ is a prefactor, which has an algebraic $T$-dependence.
Parameter $T_{\rm ES}$ is given by
\begin{equation}
T_{\rm ES} = C e^2 / \kappa \xi,
\label{TES}
\end{equation}
where $\xi$ is the (isotropic) decay length of localized electron
states, $\kappa$ is the (isotropic) dielectric constant of the
semiconductor, and $C$ is a numerical coefficient. (We measure
temperature in energy units throughout this paper.) In lightly doped
isotropic semiconductors $\kappa$ and $\xi$ are determined solely by
material parameters (binding energy of impurity states, electron
effective mass, band-structure, {\it etc.\/}). Therefore, $T_{\rm ES}$ does
not depend on the impurity concentration and Eqs.~(\ref{DC})
and~(\ref{TES}) are in this sense universal.

In contrast, in this paper we show that in strongly {\it anisotropic\/}
systems the Coulomb gap has, in general, a different functional form.
Depending on $l$ and other parameters, it may be either universal or not
[i.e., $g(\varepsilon)$ and $T_{\rm ES}$ may contain factors related to
impurity concentration]. These results and their consequences for the
VRH transport are presented in the next section.


\section{Results}
\label{Sec:Results}

We group our results according to the three cases (A, B, and C) outlined
in the Introduction.

\subsection{Single chain}

In this case, studied in Sec.~\ref{Sec:1D}, the Coulomb interaction is
not screened. However, in 1D the $1/x$-decay of the Coulomb potential is
on the borderline between the short and the long range interactions.
Consequently, most physical quantities differ from their counterparts
for the short-range (screened) interaction only by some logarithmic
factors. For example, the density of states of charge excitations
exhibits a logarithmic suppression,~\cite{Raikh}
\begin{equation}
g(\varepsilon) = \frac{g_B}{\ln (e^2 / \kappa l |\varepsilon|)}.
\label{CG_1D}
\end{equation}
In a first approximation, such a suppression can be disregraded in the
calculation of the VRH transport. Namely, one can assume that
$g(\varepsilon) = g_B = {\rm const}$. In this approximation one arrives
at the conventional Mott VRH,~\cite{Mott_book,ES75} which in 1D
coincides with the ES law of Eq.~(\ref{DC}). Let us denote by $T_{\rm
ES}^{(0)}$ the value of $T_{\rm ES}$ that one obtains neglecting the
Coulomb gap, then $T_{\rm ES}^{(0)} \sim 1 / g_B \xi_x$. Here $\xi_x$
stands for the localization length that determines the asymptotic decay
$P \propto \exp(-2 x / \xi_x)$ of the probability of tunneling of
charge-$e$ excitations over a large distance $x$. If the probability of
tunneling between nearest rods is written in the form $\exp(-2 s)$,
where $s \gg 1$, then tunelling paths with returns can be
neglected. In this case $\xi_x = l / s$. Using the expression for $s$
from Ref.~\onlinecite{Glazman_92}, we obtain
\begin{eqnarray}
 \xi_x &\sim& \frac{l}{r_s^{1/2} \ln^{3/2}({l}/{a})},
\label{zeta_1D}\\
 T_{\rm ES}^{(0)} &=& C_{1} \frac{e^2}{\kappa l} r_s^{1/2} \ln^{5/2}({l}/{a}).
\label{TES_0_1D}
\end{eqnarray}
In the last equation we absorbed numerical factors into the coefficient
$C_1 \sim 1$.

A similar expression for $T_{\rm ES}^{(0)}$ was obtained by
Nattermann~{\it et al.\/}~\cite{Nattermann_03} for the model of a
disordered Luttinger liquid with short-range interactions and weak
pinning. Our Eq.~(\ref{zeta_1D}) and (\ref{TES_0_1D}) differ from the
corresponding results of Ref.~\onlinecite{Nattermann_03} by two
logarithmic factors. One of them originates from logarithmic charging
energy of metallic rod; the other --- from the logarithm in the
tunneling action $s$, see also Refs.~\onlinecite{Glazman_92},
\onlinecite{Kane_92}, and~\onlinecite{Larkin_78}.

Once the logarithmic Coulomb gap is taken into account, the
$T$-dependence of the conductivity can still be written in the form of a
ES law [Eq.~(\ref{DC})] but $T_{\rm ES}$ becomes a function of $T$, as
follows:
\begin{equation}
T_{\rm ES} = T_{\rm ES}^{(0)} \ln
\left( \frac{e^2}{\kappa l T_{\rm ES}^{(0)}}
       \sqrt{\frac{T_{\rm ES}^{(0)}}{T}}
\right).
\label{TES_1D}
\end{equation}
Note that in the 1D case, the standard derivation~\cite{Mott_book} of
the VRH law~(\ref{DC}) overlooks the role of very resistive hops in some
exponentially rare places along the chain. A more careful approach
shows~\cite{Shante,Larkin_78} that Eq.~(\ref{DC}) and its generalization
through Eq.~(\ref{TES_1D}) are valid {\it only if the chain is
sufficiently short\/}. The quantitative criterion on the length of the
chain can be obtained following the excellent discussion in
Refs.~\onlinecite{Shante} and \onlinecite{Larkin_78}. This, however,
goes beyond the scope of the present work.

\subsection{3D systems with large impurity concentrations}

This case, formally defined by the inequality $a_\perp \ll l \ll l_s$ is
studied in Sec.~\ref{Sec:3D_dirty}. It may be realized in strongly
anisotropic CDW compounds such as KCP where the soliton length $l_s$ is
large ($10^2 a$ or so) and/or in samples where a relatively high impurity
concentration is created
intentionally~\cite{Rouziere_00,Zuppiroli_review} so that $l$ is small.
Possible non-CDW realizations include arrays of
relatively distant 1D conductors, e.g., quantum wires,
nanotubes,~\cite{Benoit_02} or polymers.~\cite{Yoon_95}

As elaborated in Sec.~\ref{Sec:Introduction}, impurities divide the
system into a collection of metallic rods. The finite 3D
concentration of highly polarizable rods results in a large dielectric
constant~\cite{Rice_72} along the $x$-axis. The Coulomb interaction is
therefore strongly anisotropic but the Coulomb gap remains parabolic,
$g(\varepsilon) \propto \varepsilon^2$, as in isotropic systems.
Tunnelling is anisotropic as well. The interchain tunnelling is
accomplished by single-electron like excitations, which do not perturb
charges on the intermediate chains along the tunneling path. In
App.~\ref{Sec:Tunneling} we estimate the corresponding transverse
localization length $\xi_\perp$ to be 
\begin{equation}
\xi_\perp = \frac{a_\perp}{\ln(e^2 / \kappa a t_\perp)},
\label{eta}
\end{equation}
where $t_\perp$ is the interchain hopping matrix element in
the tight-binding band-structure model. For the
low-$T$ VRH conductivity we again obtain the ES law with
$T_{\rm ES}$ given by
\begin{equation}
\label{DCT} T_{\rm ES} = C_{2} \frac{e^{2}}{\kappa l} \left[
\frac{a_\perp^2 \sqrt{r_s}}{\xi_\perp^2} \ln\left(\frac{l}{a_\perp}\right)
\right]^{1/3}.
\end{equation}
Here $C_2$ is another numerical factor of the order of unity. We see
that in both cases, A and B, the ES law looses its universality, because
$T_{\rm ES}$ depends on the impurity concentration $N$ through $l = 1 / N
a_\perp^2$. For a single chain (case A) this dependence originates mainly from
the dependence of the localization length $\xi_x$ on $l$. In 3D (case
B), the full effective dielectric constant and therefore, the density of
states inside the Coulomb gap depend on $N$ as well. In both cases, with
decreasing $N$ the temperature $T_{\rm ES}$ decreases, which at a fixed
temperature leads to an exponentially increasing conductivity.

In doped semiconductors similar violations of the universality of
Eq.~(\ref{DC}) are known to occur near the metal-insulator transition.
In that case, however, $T_{\rm ES}$ has an {\it opposite\/} dependence on
$N$. In particular, $T_{\rm ES}$ vanishes when $N$ grows and reaches the
critical concentration.~\cite{ES84} Similarly, previous theories
of the VRH transport in strongly anisotropic systems dealt with gapped,
semiconductor-like materials (commensurate CDW) where impurities
provided carriers,~\cite{Kivelson,Li_93} so that the conductivity was found to
{\it grow\/} with the impurity concentration. In contrast, our work is
devoted to systems, which are metallic (sliding) in the absence of
impurities. Therefore, decrease of $T_{\rm ES}$ with decreasing $N$ seems
natural.~\cite{Kivelson,Li_93}

At high temperatures, the conductivity is due to the nearest-neighbor
hopping. Its $T$-dependence is of activated type,
\begin{equation}
\label{DCROD}
\sigma = \sigma_A \exp(-E_A / T), \quad l < l_s,
\end{equation}
with the activation energy
\begin{equation}
\label{E_A}
E_A \sim \frac{e^2}{\kappa l}
\end{equation}
and the prefactor $\sigma_A$ proportional to the
probability of tunneling between adjacent rods. 

\subsection{3D systems with small impurity concentration}

\begin{figure}
\includegraphics[width=2.7in]{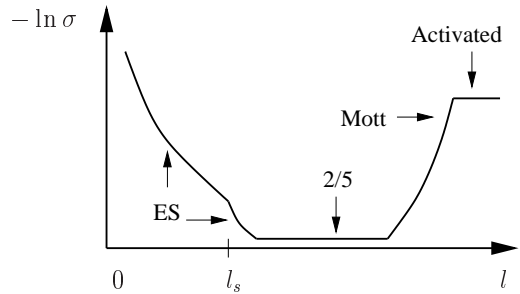}
\vspace{0.1in}
\caption{ \label{sigma3D_l} Logarithm of the resistivity as a
function of the average inter-impurity distance $l = 1 / N a_\perp^2$ at
a fixed temperature $T \ll W$. The ES, 2/5, Mott, and
activation laws succeed
each other with growing $l$.}
\end{figure}

As impurity concentration decreases and $l$ becomes larger than $l_s$, a
number of dramatic changes appear in all key quantities, such as the
density of states, the localization length, and the effective dielectric
constant. For example, as we discuss in Sec.~\ref{Sec:3D_clean}, the
dielectric constant starts to increase exponentially with $l$ because
the polarizability of the crystal with $l > l_s$ becomes limited not by
the length $l$ of individual chain segments but by the exponentially
large length of Larkin domains. The soaring dielectric constant causes a
rapid drop of the ES parameter $T_{\rm ES}$. In turn, this causes a collapse
of the low-temperature resistivity in a narrow interval $l_s \lesssim l
\lesssim l_s \ln(W / T)$ (see the descending branch of the curve in
Fig.~\ref{sigma3D_l}). Until this point, the notion that our system is
opposite to the conventional semiconductors, so that purer samples have
higher conductivities, seem to be working.

Once $l$ exceeds $l_s \ln(W / T)$, the Larkin
length $L_x$ can be treated as effectively infinite. The VRH now
involves hops between low-energy states separated by distances shorter
than $L_x$. On such distances, the dispersion of the dielectric function
becomes important. Each pair of low-energy charge excitations localized
on their respective impurity clusters interacts via a strongly
anisotropic electrostatic potential, which is not exponentially small
only if the vector that connects the two charges is nearly parallel to
the chain direction. Such an unusual interaction leads to a Coulomb gap
that is linear in energy and independent of $N$, unlike the previous
case (case B, $l \ll l_s$), where the Coulomb gap is quadratic and
$N$-dependent. Another difference from the case B is that the
localization length $\xi_x$ for the tunnelling in the chain direction is
also independent of $N$,
\begin{equation}
\xi_x \sim \frac{l_s}{\sqrt{r_s}},
\label{zeta_clean}
\end{equation}
see Sec.~\ref{Sec:3D_clean} and App.~\ref{Sec:Tunneling}. This
leads to a novel 2/5-law for the VRH conductivity,
\begin{equation}
\sigma = \sigma_0 \exp[-(T_{1}/T)^{2/5}],
\label{DC1}
\end{equation}
where parameter $T_1$, given by
\begin{equation}
\label{DCT1}
T_{1} = C_{3} \frac{e^2 r_s^{1/4}}{\kappa l_s}\frac{a_\perp}{\xi_\perp},
\end{equation}
does not depend on $l$ and is, in this sense, universal. Here $C_3$ is yet
another numerical coefficient of the order of unity. The 2/5-law shows
up as an intermediate resistivity plateau in Fig.~\ref{sigma3D_l}. This
universal law for quasi-1D systems with $l \gg l_s$ is an analog of
the universal ES law in isotropic systems. 

We show in Sec.~\ref{Sec:3D_clean} that the Coulomb gap affects mainly a
finite energy interval $|\varepsilon| \lesssim \Delta$, where $\Delta
\propto g_B$ can be called the Coulomb gap width. At larger energies,
the density of states of charge excitations coincides with the bare one,
$g(\varepsilon) \simeq g_B$. Since $g_B$ is generated by impurity
clusters whose concentration diminishes with growing $l$, both $g_B$ and
$\Delta$ decrease with $l$. Eventually, the Coulomb gap becomes more
narrow than the range of energies around the Fermi level responsible for
hopping at given $T$. At this point the Coulomb gap can be neglected and the
2/5-law is replaced by the conventional Mott law,
\begin{equation}
\sigma = \sigma_0 \exp[-(T_{M}/T)^{1/4}],
\label{DCM}
\end{equation}
where $T_M = C_4 / g_{B} \xi_x \xi_\perp^2$, $C_4 \sim 1$. As $l$
increases further, $T$ being fixed, $\sigma$ decreases because of
diminishing $g_B$. This gives rise to the ascending branch in
Fig.~\ref{sigma3D_l}. At such $l$, the electron crystal behaves similar
to a gapped insulator where a lower impurity concentration
corresponds to a lower carrier density, and thus, to a higher
resistivity.

As $l$ continues to grow, at some point the Mott VRH crosses over to the
nearest-neigbor hopping and shortly after it becomes smaller than the
conductivity due to thermally activated free solitons,
\begin{equation}
\label{DCSOL}
                  \sigma = \sigma_A \exp(-W / T).
\end{equation}
At even larger $l$, $\sigma$ ceases to depend on $l$, and so the
impurity concentration, see Fig.~\ref{sigma3D_l}. Note that the
activation energies $W \sim e^2 / \kappa l_s$ [Eq.~(\ref{DCSOL})] and
$E_A \sim e^2 / \kappa l$ [Eq.~(\ref{E_A})] in the cases A and B,
respectively, smoothly match at $l \sim l_s$.

\subsection{Summary of the regimes}

\begin{figure}
\includegraphics[width=2.2in]{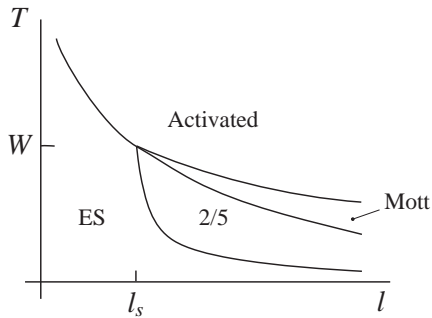}
\vspace{0.1in}
\caption{\label{PD_3D} Summary diagram for the transport regimes in a 3D
system. Domains of validity of ES [Eq.~(\ref{DC})], Mott
[Eq.~(\ref{DCM})], activated [Eqs.~(\ref{DCROD}) and~(\ref{DCSOL})] and
2/5 [Eq.~(\ref{DC1})] laws are shown.}
\end{figure}

The rich behavior of the conductivity as a function of $l$ and $T$ is
summarized in the form of a regime diagram in Fig.~\ref{PD_3D}. The
novel $2/5$-law applies in a broad range of $l$ and $T$ between the ES
and the Mott laws. 

A convenient way to keep track of all the VRH exponents derived in this
paper is provided by Eq.~(\ref{VRH_exponent}) below. We would like to
present it in a somewhat more general form, motivated by the following
physical reasoning.

The diagram of Fig.~\ref{PD_3D} is obtained under the assumption that
the tunneling in the transverse direction is not negligible
($\xi_\perp$ is not too small), so that the VRH has a 3D
character. However, if the conducting chains are relatively distant from
each other either along $y$ or $z$-direction or both, this condition may
be violated. Examples of such systems are artificial arrays of quantum
wires~\cite{Mani_92} and carbon nanotubes.~\cite{deHeer_95} In those
systems, the 3D hopping is pushed to very lower temperatures, while at
intermediate $T$ the hopping can be either one or two-dimensional.
Generalizing the standard derivation of the VRH
transport~\cite{Mott_book,ES84} to the
$d$-dimensional hopping and a power-law density of states $g(\varepsilon)
\propto \varepsilon^\mu$, one obtains the conductivity in the form
\begin{equation}
\sigma = \sigma_0 \exp \left[
-\left({T_{\rm VRH}}/{T}\right)^{\lambda}\right],\quad
\lambda = \frac{\mu + 1}{\mu + d + 1}.
\label{VRH_exponent}
\end{equation}
For $d = 3$ one recovers all the regimes discussed prior in this Section
(Mott, $2/5$, and ES laws) by setting $\mu$ successively to $0$, $1$,
and $2$, according to the physical situation. For the sake of
completeness, the exponents for other $d$'s in the same situations are
summarized in Table~\ref{Table_VRH}. Inclusion of all such regimes would
transform Fig.~\ref{PD_3D} into a more complicated diagram, but would
not change its general structure, so it will not be shown here or
discussed further below.

\begin{table}
\caption{\label{Table_VRH} The exponents $\lambda$ of VRH
conductivity [Eq.~(\ref{VRH_exponent})] in the cases
of 3D, 2D and 1D tunneling and a power-law dependent density of states
$g(\varepsilon)$ that arises due to 3D Coulomb interactions. 1D tunneling
corresponds to $\xi_\perp \to 0$.}
\begin{ruledtabular}
\begin{tabular}{cccc}
Tunneling & $g = {\rm const}$ & $g \propto |\varepsilon|$ & $g \propto \varepsilon^2$\\
\hline
3D        & 1/4               &      2/5               & 1/2                    \\
2D        & 1/3               &      1/2               & 3/5                    \\
1D        & 1/2               &      2/3               & 3/4                    
\end{tabular}
\end{ruledtabular}
\end{table}


\section{1D system}
\label{Sec:1D}

In the case of a single chain of electrons on a uniform positive
background (Fig.~\ref{Pinned_CDW}a) impurities divide the 1D electron
crystal in separate pieces, which behave as metallic rods. The rod
lengths $x$ are distributed randomly around the average value $l$.
Therefore, the background charge of a given rod, $Q = -e x / a$, is a
random number. It can be written as $Q = -e(n + \nu)$, where $n$ is an
integer and $\nu$ is a number uniformly distributed in the interval
$-1/2 < \nu < 1/2$.

In the ground state of the system each rod contains
an integer number $n_r$ of electrons, so that the rod has the total
charge of $q = e (n_r - n - \nu)$. To find $n$ we use the fact that the
Coulomb self-energy of the rod is equal to $q^2 / 2 C_r$ where $C_r =
\kappa x / [2 \ln(x/a)]$ is the capacitance of this rod. On the other
hand, the interaction of different rods does not contain the large
logarithm $\ln(x/a)$, and can be neglected in the first approximation.
Thus, the minimization of the total energy of the system amounts to
minimizing the self-energies of the rods. One can show then that, in the
ground state, $n_r = n$, so that the charges of the rods are uniformly
distributed in the interval $-e/2 < q < e/2$. Indeed, if this is not true
and the charge of the given rod is $q > e/2$, then, by charge conjugation
symmetry, there should exist another rod with the same length and the
opposite charge $-q$. Transferring an electron from the first rod to the
second one we lower the total energy and make the absolute values of
both charges smaller than $e/2$.

Below we use $q = -e \nu$ and call rods with $0 < \nu < 1/2$ empty and with
$-1/2 < \nu <0$ occupied. We define the energy of an empty state,
$\varepsilon(x,\nu)$, as the minimum work necessary to bring to it an
electron from a distant pure 1D electron crystal with the same average
linear density of electrons
\begin{eqnarray}
\varepsilon(x,\nu) &=& \frac{e^2}{\kappa x}
\ln\left(\frac{x}{a}\right) [(1 - \nu)^2 - \nu^2]
\nonumber\\
&=& \frac{e^2}{\kappa x}
\ln\left(\frac{x}{a}\right) (1 - 2\nu),
\label{nrg_elec1D}
\end{eqnarray}
This energy is positive and vanishes only at $\nu=1/2$. On the other
hand, the energy of an occupied state is defined as minus the maximum
work necessary to extract electron from this rod to the same distant
pure 1D electron crystal. In this case the final result is identical to
Eq.~(\ref{nrg_elec1D}) with $\nu \to -\nu$. Apparently, states with
$|\nu| = 1/2$ are exactly at the Fermi level, which we take as the
energy reference point (it coincides with the electron chemical
potential of a pure 1D crystal). The low-energy states relevant to VRH
transport have $|\nu| - 1/2 \ll 1$.

Now we can calculate the density of such states. Taking
into account the fact that this rod length $x$ is distributed
according to Poisson statistics, the disorder-averaged
bare density of states can be written as
\begin{equation}
\label{DOS1D_def}
g_B(E)= \frac{1}{l}\int\limits_0^{{1 / 2}}{d\nu}
\int\limits_0^\infty \frac{dx}{l}
\exp \left( -\frac{x}{l} \right)
\delta \left(E - \varepsilon(x,\nu) \right).
\end{equation}
With a logarithmic accuracy, we can replace $\ln(x/a)$ by $\ln(l/a)$.
Then we use Eqs.~(\ref{nrg_elec1D}) and (\ref{DOS1D_def}) to find
\begin{equation}
\label{DOS1D_expl}
g_B(\varepsilon)= \frac{1}{2 \varepsilon_0 l}
\left[1- \left( 1+\frac{\varepsilon_0}{\varepsilon} \right)
\exp \left(-\frac{\varepsilon_0}{\varepsilon} \right) \right],
\end{equation}
where $\varepsilon_0 = e^2 \ln(l/a)/ \kappa l$. We warn the reader that
one should not attribute much significance to the predictions of the
above formula in the region of high energies, $\varepsilon \gtrsim
\varepsilon_0$, where excitations with charges larger than $e$ will also
contribute to various physical processes. On the other hand, close to
the Fermi level, for $\varepsilon \ll \varepsilon_0$, only charge-$e$
excitations are important, in which case Eq.~(\ref{DOS1D_expl}) is
fully adequate, while $g_B$ is nearly constant,
\begin{equation}
\label{g_0}
                g_B \simeq \frac{\kappa}{2 e^2 \ln (l / a)}.
\end{equation}
For the calculation of the low-$T$ transport, only this constant value
is needed.

As mentioned in Sec.~\ref{Sec:Results}, in 1D the $1 / x$-Coulomb
interaction creates only marginal effects on the conductivity. If we
neglect them, in the first approximation, then standard Mott's
argument~\cite{Mott_book} leads to the VRH that obeys the $T^{1/2}$-law,
Eq.~(\ref{DC}), with $T_{\rm ES} = T_{M} = C_{1}/g_B\xi_x$, where
$\xi_x$ is localization length for tunnelling between distant rods. The
value of $\xi_x$ is obtained from the following considerations.
Tunneling through an impurity that separates two adjacent rods can be
viewed as a process in imaginary time that consists of the following
sequence of events.~\cite{Larkin_78} A unit charge assembles into a
compact soliton just before the impurity in one rod, tunnels through the
impurity, and finally spreads uniformly over the other rod. We assume
that the chain is not screened by external metallic gates. Then the
tunneling probability can be written in the form $\exp(-2s)$, where the
$s$ is the dimensionless action~\cite{Glazman_92} $s \sim r_s^{1/2}
\ln^{3/2}(l/a)$. Since $s \gg 1$, tunneling paths with returns can be
neglected. Therefore, for electron tunnelling over distances $x \gg l$
the action $s$ should be multiplied by the number of impurities passed.
The average number of such impurities is equal to $x / l$, which yields
the total tunnelling probability $P \propto \exp(-2 x / \xi_x)$, with
$\xi_x$ given by Eq.~(\ref{zeta_1D}). With a logarithmic accuracy, the
effect of the Coulomb gap is to replace $g_B$ by $g(\varepsilon)$
evaluated at the characteristic hopping energy $\varepsilon =
\sqrt{T_{\rm ES}^{(0)} T}$, where $T_{\rm ES}^{(0)}$ is defined by
Eq.~(\ref{TES_0_1D}). The latter result follows from Eq.~(\ref{g_0}) and
(\ref{zeta_1D}). The final expression for the parameter $T_{\rm ES}$ is
given by Eq.~(\ref{TES_1D}). We briefly note that in a long enough chain
clusters of atypically densely spaced impurities may exist. Tunneling
through such segments would cost a higher tunneling action and therefore
the overall conductivity would be suppressed. In this paper we assume
that the chain is sufficiently short so that these rare clusters can be
neglected, see a comment after Eq.~(\ref{TES_1D}).

Formulas~(\ref{TES_0_1D}) and (\ref{TES_1D}) indicate that $T_{\rm
ES}$ goes down as impurity concentration $N = 1 / l$ decreases. This
provides a gradual crossover to the metallic behavior in a pure 1D
system.


\section{3D system with a large impurity concentration}
\label{Sec:3D_dirty}

In this section we consider a quasi-1D system made of parallel chains
which form a periodic array in the transverse directions. The chains are
pinned by impurity centers, which divide them into metallic rods of
average length $l = 1 / N a_\perp^2$, where $a_\perp^2$ is the cross-sectional area
per chain. The finite-size electron crystals in each rod are either
compressed or stretched to accommodate an integral number of electrons,
as in the case of a single chain. In this section we assume that for a
typical rod, the energy of its longitudinal deformation is smaller than
the energy of its transverse coupling to rods on the neighboring chains.
This can be the case when the periodic potential created by the
neighboring chains is diminished because $a_\perp$ is larger than $a$, and/or
when the impurity concentration is large enough. Formally, the
inequality $a_\perp \ll l \ll l_s$ needs to be satisfied, where $l_s$ is the
soliton length [Eq.~(\ref{l_s_def})]. Since metallic rods now completely
fill the 3D space (see Fig.~\ref{Pinned_CDW}b), they modify the
dielectric constant of the system. As in the interrupted-strand
model,~\cite{Rice_72} the dielectric constant is anisotropic. Along the
chain direction it has the value of
\begin{equation}
\kappa_{x} = \kappa [1 + C_5 (l / r_D)^2],
\label{kappa_x_dirty}
\end{equation}
where $r_D \sim a_\perp$ is the screening length, see Apps.~\ref{Sec:Elasticity},
and $C_5 \sim 1$ is a numerical constant.~\cite{Commend_on_C_5}
Transverse dielectric constants are unaffected,
$\kappa_{y}=\kappa_{z}=\kappa$. At large $l$, $\kappa_x$ greatly exceeds
$\kappa$, which leads to an anisotropic Coulomb interaction in the
form~\cite{Landau}
\begin{equation}
\label{phi_ES}
U(r) = \frac{e^2}{\kappa \sqrt{x^2 + (\kappa_x / \kappa) r_\perp^{2}}},
\end{equation}
where $r_\perp^{2} = y^2 + z^2$. In spite of the large dielectric constant,
the Coulomb interaction is long-range and thus creates a Coulomb gap.
Using the standard ES argument,~\cite{ES84,Efros_76} the following
density of states of charge excitations is obtained
\begin{equation}
\label{CG_ES}
g(\varepsilon) =\frac{3}{\pi} \frac{\kappa^{2}\kappa_x}{e^6}
 \varepsilon^2,
\end{equation}
It differs from the conventional formula for an isotropic medium only by
the presence of $\kappa_x$ instead of $\kappa$. In order to calculate
the VRH conductivity we still have to discuss the tunnelling
probability. It is important that Eq.~(\ref{phi_ES}) holds only at $x
\gg l$. The interaction between charge fluctuations on the same rod is
short-range due to screening by neighboring
chains.~\cite{Fogler_unpublished} Tunnelling along the $x$-axis takes
place similarly to the case of a single chain, but screening of the
Coulomb interaction leads to smaller action of the order of $s \sim
\sqrt{r_s} \ln(l/a)$. Therefore, for the localization length in the
$x$-direction we get
\begin{equation}
\label{zeta_dirty}
\xi_x = \frac{l}{s} \sim \frac{l}{\sqrt{r_s} \ln(l/a)}.
\end{equation}
The tunneling in the $y$- and $z$-directions is accomplished by
single-electron like excitations. The probability of tunneling decays
exponentially, $P \propto \exp(-2r_\perp/\xi_\perp)$, at large transverse
distances $r_\perp$. Here $\xi_\perp$ is the transverse localization
length given by Eq.~(\ref{eta}) and discussed in more detail in
App.~\ref{Sec:Tunneling}. We assume that $\xi_\perp$ is not vanishingly
small compared to $\xi_x$, in which case the VRH has a 3D character and
can be calculated by the percolation approach (see
Ref.~\onlinecite{ES84} and references therein).~\cite{ES_law_1D_hopping}
This calculation differs from the isotropic case by the replacement of
the isotropic dielectric constant $\kappa$ and the isotropic
localization length $\xi$ by their geometric averages over the three
spatial directions:
\begin{equation}
T_{\rm ES} = C_3 \frac{e^2}{(\kappa^2 \kappa_x)^{1/3}}.
\frac{1}{(\xi_x\xi_\perp^2)^{1/3}}
\label{T_ES_anisotropic}
\end{equation}
With the help of Eqs.~(\ref{kappa_x_dirty}) and (\ref{zeta_dirty}) the
expression for $T_{\rm ES}$ reduces to Eq.~(\ref{DCT}), where the screening
length of the electron crystal has been taken as $r_D \sim a_\perp$ and
numerical coefficients absorbed in $C_3$.


\section{3D system with a small impurity concentration}
\label{Sec:3D_clean}
 
In this section we study the crystal pinned by impurities with a low
concentration $N$ so that the condition $l \gg l_s$ is satisfied.
As discussed in Sec.~\ref{Sec:Introduction} all key quantities --- density
of states, localization length, the screening of Coulomb
interactions --- undergo dramatic changes compared to the case of high
impurity concentration. We address such changes in the three separate
subsections below.


\subsection{Pinning of a quasi-1D crystal by strong dilute impurities}
\label{Sec:Pinning}

In this subsection we study the ground state structure and screening
properties of the crystal with low impurity concentration.

We start by reviewing the physical meaning of $l_s$ given by
Eq.~(\ref{l_s_def}), in which $\alpha = Y_\perp / Y_x$ is the anisotropy
parameter, and $Y_x$, $Y_\perp$ parametrize the energy of an elastic
distortion of the crystal,
\begin{equation}
     E^{\rm el} = \frac12 \int d^3 r
            [Y_x (\partial_x u)^2 + Y_\perp (\nabla_\perp u)^2].
\label{E_elastic}
\end{equation}
As shown in App.~\ref{Sec:Elasticity}, at large $r_s$ the longitudinal
elastic modulus $Y_x$ is dominated by Coulomb effects,
\begin{equation}
                    Y_x \sim e^2 / \kappa a^2 a_\perp^2.
\label{Y_x}
\end{equation}
The transverse modulus $Y_\perp$ can be substantially smaller than $Y_x$
even when the ratio $a_\perp / a$ is only modestly large. For Coulomb
interaction, $Y_{\perp} \propto \exp(-2\pi a_\perp/a)$.

The energy $E^{\rm el}$ in Eq.~(\ref{E_elastic}) is essentially the
short-range part of the Coulomb energy. The total energy also includes
the long-range Coulomb part (see App.~\ref{Sec:Elasticity}) and the
pinning part (see App.~\ref{Sec:Collective} and below). Strictly
speaking, Eq.~(\ref{E_elastic}) is valid only for small gradients of the
elastic displacement field $u({\bf r})$; however, it can be used for
order-of-magnitude estimates down to microscopic scales $r_\perp \sim a_\perp$
and $x \sim l_s$. In this manner, one can derive formula~(\ref{l_s_def}) by
minimizing $E^{\rm el}$ under the condition that $u$ changes from $0$ to
$a$ over a segment of length $l_s$ on a single chain. For more details,
see Refs.~\onlinecite{Brazovskii_91},~\onlinecite{Larkin_94}, and
App.~\ref{Sec:Elasticity}.

The inequality $l \gg l_s$ imposes the upper limit on the
impurity density:
\begin{equation}
                    N \ll \sqrt{\alpha} / a_\perp^3.
\label{Condition_on_N}
\end{equation}
Below we show that at such $N$ the ground state of the electron crystal
is determined by an interplay of individual and collective
pinning.~\cite{Abe}

Without impurities the crystal would have a perfect periodicity and
long-range 3D order. Impurities cause elastic distortions of the
lattice. The strongest distortions, of dipolar type, are localized in
the vicinity of impurities, see Sec.~\ref{Sec:Introduction} and
Fig.~\ref{Pinned_CDW}c. Such dipoles have a characteristic size $l_s$
(same as free solitons), are well separated from each other, and occupy
only a small fraction of the space. Their creation is advantageous
because the associated energy cost (elastic plus Coulomb) is of the
order of $W \sim e^2 / \kappa l_s$ per impurity, whereas the energy gain
is a much larger electron-impurity interaction energy $-e^2 / \kappa a$.
This is the essence of individual (strong) pinning phenomenon, which
provides the dominant part of the pinning energy density. The collective
(weak) pinning results from interaction between the dipoles. Let us
demonstrate that such interaction cannot be neglected at sufficiently
large length scales. By solving the elasticity theory equations
(generalized to include the Coulomb interactions), it can be
shown~\cite{Brazovskii_91} that a dipolar distortion centered at a point $(x_i,
r_{\perp i})$ has long-range tails that decay rather slowly with distance, $u
\sim A_i a r_D / \sqrt{\alpha} |x - x_i|$. This displacement is confined
mainly within a paraboloid $|r_\perp - r_{\perp i}|^2 \lesssim \sqrt{\alpha} r_D
|x - x_i|$. Note that the segment $0 < x < x_{\rm min} \equiv
a_\perp (l / \sqrt{\alpha} r_D)^{1/2}$ of the paraboloid
\begin{equation}
                    r_\perp^2 = \sqrt{\alpha} r_D |x|
\label{paraboloid}
\end{equation}
contains on average one impurity. Parameter $r_D \sim a_\perp$,
which we already encountered in Sec.~\ref{Sec:3D_dirty}, has
the meaning of the screening length. It is related to the longitudinal
elastic modulus $Y_x$ as follows:
\begin{equation}
r_D^2 = \frac{\kappa}{4 \pi e^2}  a^2 a_\perp^4 Y_x,
\label{r_D}
\end{equation}
see App.~\ref{Sec:Elasticity}.

If, in the first approximation, we choose to neglect the interaction
among the dipoles, then we should simply add their far elastic fields
treating the amplitudes $-1 \lesssim A_i \lesssim 1$ as random
variables. We immediately discover the logarithmic growth of $u$ with
distance,
\begin{eqnarray}
\langle [\bar{u}(x, 0) - \bar{u}(0, 0)]^2\rangle
&\sim& N \int\limits_{x_{\rm min}}^x d x^\prime
\int\limits_0^{\sqrt{\alpha} r_D x^\prime}
d r_\perp^2 \left(\frac{a r_D}{\sqrt{\alpha} x^\prime}\right)^2
\nonumber\\
       &=& \left(\frac{a^2}{C_6}\right) \left(\frac{l_s}{l}\right)
           \ln \left(\frac{x}{x_{\rm min}}\right),
\label{u_squared}
\end{eqnarray}
where the bar over $u$ indicates that we refer to the value of $u$ away
from the immediate vicinity of a dipole and $\langle\ldots\rangle$
stands for disorder averaging.

The logarithmic growth of the elastic displacements was previously
derived for the model of {\it weak\/} pinning centers in early
works~\cite{Bergman_77,Efetov_77} on the subject. In those calculations
the numerical coefficient $C_6$ is large and is inversely proportional
to the impurity strength. In our case $C_6 \sim 1$. Apart from that,
Eq.~(\ref{u_squared}) demonstrates that the case of {\it strong\/}
pinning centers is essentially similar. Therefore, as customary for weak
pinning models we define the longitudinal $L_x$ and transverse $L_\perp$
Larkin lengths as the lengthscales where $\langle\Delta u^2\rangle
\equiv \langle [u({\bf r}) - u(0)]^2\rangle \sim a^2$. From
Eqs.~(\ref{paraboloid}) and (\ref{u_squared}) we obtain
\begin{equation}
\label{larkin_l}
L_x = x_{\rm min} \exp(C_6 l / l_s),\quad
L_\perp = r_\perp^{\rm min} \exp(C_6 l / 2 l_s),
\end{equation}
where $r_\perp^{\rm min} = (\sqrt{\alpha} r_D x_{\rm min})^{1/2}$.

Alternative derivation of Eqs.~(\ref{u_squared}) and (\ref{larkin_l})
based on energy estimates is given in App.~\ref{Sec:Collective}. It
elucidates that the slow logarithmic growth of $\langle\Delta
u^2\rangle$ is rooted in the important role of long-range Coulomb
interaction in the elastic response of a quasi-1D crystal. An isotropic
electronic crystal adjusts to pinning centers primarily by means of
shear deformations~\cite{Fogler_00} that do not cost much Coulomb
energy. As a result, in the isotropic crystal $\Delta u$ grows
algebraically with distance. In contrast, in quasi-1D crystals and CDW,
where elastic dispacement is a scalar (electrons move only along the
chains), no separate shear deformations exist. The build-up of the
Coulomb energy that accompanies longitudinal compressions translates
into an exceptionally large rigidity of the electron lattice and
exponentially large $L_x$ and $L_\perp$.

At distances exceeding the Larkin lengths the dipoles can no
longer be treated as independent. Indeed, the energy cost $E_s$ of a
given dipole is determined by the minimal distance by which the crystal
has to distort to align an electron with the impurity position.
Therefore, just like the energy of a rod in the previous sections, $E_s$
has a periodic dependence on $\nu = \{(x_i - \bar{u}) / a\}$, where
$\{\ldots\}$ denotes the fractional part. $E_s$
vanishes at $\nu = 0$ and reaches a maximum value of $\sim W$ at $\nu =
\pm 1/2$. Therefore, as soon as the cumulative effect of other dipoles
attempts to elevate $|\nu|$ above $1/2$, a $2\pi$-phase slip should
occur to adjust $E_s(\bar{u})$ to a lower value. The overall effect of
such adjustments is to enhance the pinning energy. This
additional energy gain can be viewed as the {\it collective pinning\/}
effect. Using standard arguments (see App.~\ref{Sec:Collective}), we
relate the extra pinning energy density to the Larkin length
\begin{equation}
{\cal E}_{\rm pin} \sim -W / a_\perp^2 L_x \sim -Y_\perp (a / L_\perp)^2.
\label{epsilon_pin}
\end{equation}
We will now use this result to estimate the asymptotic value
\begin{equation}
\kappa_x \equiv \epsilon(q_x \to 0, q_\perp = 0)
\label{kappa_x_def}
\end{equation}
of the longitudinal component of the dielectric function.

Without impurities, the dielectric function has the form
\begin{equation}
\epsilon({\bf q}) = {\kappa} + \frac{q_x^2}{q^2}
\frac{\kappa r_D^{-2}}{q_x^2 + \alpha q_\perp^2},
\label{epsilon_q}
\end{equation}
see App.~\ref{Sec:Elasticity} and, e.g.,
Ref.~\onlinecite{Brazovskii_93}. Based on previous
work~\cite{Lee_74,Efetov_77,Gruner} we assume that a reasonable
description of dielectric screening in a system with impurities is
obtained if we replace the random distrubution of pinning centers by a
commensurate pinning with the same ${\cal E}_{\rm pin}$. In this case we
can use the concept of the dielectric function even for the disordered
system. To derive the modified expression for the dielectric function,
one can add the term ${\cal E}_{\rm pin} (u / a)^2$ to the right-hand
side of Eq.~(\ref{H_0}) in App.~\ref{Sec:Elasticity} and repeat the
steps outlined thence. It is easy to see that the net effect of
(commensurate) pinning is to augment the combination $q_x^2 + \alpha
q_\perp^2$ (proportional to the elastic resoring force) by the term
$-{\cal E}_{\rm pin} / Y_\perp a^2$, which comes from the additional
restoring force due to impurities. In this manner we obtain
\begin{eqnarray}
\epsilon({\bf q}) &=& {\kappa} + \frac{q_x^2}{q^2}
\frac{\kappa r_D^{-2}}{q_x^2 + \alpha (q_\perp^2 + L_\perp^{-2})},
\label{epsilon_q_pinned}\\
\kappa_x &=& \frac{\kappa}{\alpha} \left(\frac{L_\perp}{r_D}\right)^2
          = \frac{\kappa}{\sqrt{\alpha}} \frac{L_x}{r_D}
         \sim \exp \left(C_6 \frac{l}{l_s}\right).
\label{kappa_x}
\end{eqnarray}
However, for our purposes a cruder approximation will be sufficient.
Namely, we can assume that at distances shorter than the Larkin length,
the system screens as though it is free of impurities,
Eq.~(\ref{epsilon_q}); at distances larger than the Larkin length, the
dielectric function is replaced by a constant, Eq.~(\ref{kappa_x}). In
the following subsection we will use Eqs.~(\ref{epsilon_q}) and
(\ref{kappa_x}) to derive the functional form of the Coulomb gap in the
regime $l \gg l_s$.


\subsection{Bare density of states and the Coulomb gap at low
impurity concentration}
\label{Sec:Clusters}

In order to describe the low-$T$ transport at low impurity
concentration, $l \gg l_s$, we need to determine the origin of
low-energy charge excitations in this regime. This poses a conceptual
problem. Indeed, such excitations do not exist in the bulk (away from
impurities) where the creation energy of charge-$e$ excitations is
bounded from below by the energy $W$ of a $2\pi$-soliton. At first
glance, the impurities do not help either. As mentioned in
Sec.~\ref{Sec:Introduction}, near isolated impurities there is an energy
gap for charge excitations, which is not much smaller than $W$. This is
because a single impurity appreciably disturbs the crystal only within
the region of length $l_s$. The disturbance is electrically neutral
(dipolar) in the ground state. Creation of a charge-$e$ excitation near
such an impurity requires an energy of the order of $e^2 / \kappa l_s
\sim W$. Let us now show that charge excitations of arbitrary low
energies nevertheless exist. They come from impurity clusters. Each
cluster is a group of a few impurities spaced by distances of the order
of $l_s$ or smaller. (It can be viewed as a microscopic inclusion of the
$l \lesssim l_s$ phase.) Below we demonstrate that the clusters provide
the bare density of states at zero energy, which decreases with $l$ no
faster than a power-law,
\begin{equation}
g_B = \frac{\kappa}{e^2 a_\perp^2} \left(\frac{l_s}{l}\right)^{\beta + 2},
\label{g_B_clusters}
\end{equation}
where exponent $\beta$ is of the order of unity and is independent of $l$.
The calculation of $\beta$ has to be done numerically, which we leave
for future work.

\begin{figure}
\includegraphics[width=3.7in]{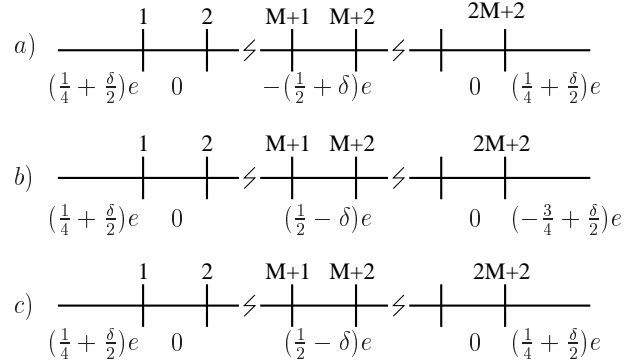}
\caption{ \label{Fig:Cluster} An example of an impurity cluster that
provides low-energy charge excitations at $l \gg l_s$. (a)
Distribution of charges in the ground state. For simplicity,
we chose $\delta_L = \delta / 2$. The cluster is electrically
neutral. (b) Competing neutral state. (c) Excited state obtained by
increasing the charge of the central rod by one unit.
}
\end{figure}

To prove that Eq.~(\ref{g_B_clusters}) gives the lower bound on $g_B$,
consider the configuration of $2 M + 2$ impurities shown in
Fig.~\ref{Fig:Cluster}. The impurities define a cluster of $2 M + 1$
short rods, each of approximately the same length $c \ll l_s$. The
cluster is flanked by two semi-infinite segments at the ends. We assume
that $M$ is sufficiently large so that $L = M c$ is much greater than
$l_s$. Suppose that the length of the central rod in units of $a$ is
close to a half-integer, so that the charge of this rod is restricted to
the set of values $q = (-1 / 2 - \delta + n) e$, where $n$ is an integer
and $0 < \delta \ll 1$. For the low energy states we only need to
consider two possibilities, $q = (-1/2 - \delta) e$ and $(1/2 - \delta)
e$. The lengths of the other short rods in our construction are chosen
to be close to integer multiples of $a$. Then those rods can be
considered charge neutral. Finally, we assume that the position
of the leftmost impurity restricts the charge of the left semi-infinite
cluster to $1/4 + \delta_L + n_L$, where $|\delta_L| \ll 1$ and $n_L$ is
another integer. Under these assumptions, the charge of the right
semi-infinite segment is fixed to the set $1/4 + \delta - \delta_L +
n_R$.

One possible candidate for the ground state is the charge configuration
shown in Fig.~\ref{Fig:Cluster}a. It is overall neutral, has the charge
of the central rod equal to $q = (-1 / 2 - \delta) e$, and the total
energy
\begin{eqnarray}
E_a &=& \frac{2}{L \kappa}\frac{e}{4}\left(-\frac{e}{2}\right)
+ \frac{1}{2 L \kappa}\left(\frac{e}{4}\right)^2
+ \frac{e^2 (1/2 + \delta)^2}{2 C_r} + 2 W_{1/4}
\nonumber\\
    &=& -\frac{7}{32} \frac{e^2}{L \kappa}
+ \frac{e^2}{2 C_r} (1/2 + \delta)^2 + 2 W_{1/4}.
\label{E_a}
\end{eqnarray}
Here we made use of the conditions $L \gg l_s$, $\delta \ll 1$.
$W_{1/4}$ denotes the self-energies of the semi-infinite segments (which
contain ``$1/4$-solitons''), and $C_r \sim \kappa c$ is the
self-capacitance of the central rod. If the configuration shown in
Fig.~\ref{Fig:Cluster}a is indeed a ground state, it has an unusual
feature that the charge of the central rod exceeds 1/2 by the absolute
value, contrary to the arguments in Sec.~\ref{Sec:1D}. To establish the
conditions for the cluster to have such a ground state, we need to
compare $E_a$ with energies of other possible states. Of those, some are
neutral and some are charged. One competing neutral state is
shown in Fig.~\ref{Fig:Cluster}b. It has energy
\begin{equation}
E_b = -\frac{11}{32} \frac{e^2}{L \kappa}
+ \frac{e^2}{2 C_r} (1/2 - \delta)^2 + W_{1/4} + W_{3/4},
\label{E_b}
\end{equation}
where $W_{3/4}$ is the energy of the ``$3/4$-soliton'' on the right of
the cluster in Fig.~\ref{Fig:Cluster}b. For the energy difference we
have
\begin{equation}
E_b - E_a = \frac{e^2}{C_r} \delta
- \frac{1}{8} \frac{e^2}{L \kappa} + W_{3/4} - W_{1/4},
\label{E_b_E_a}
\end{equation}
so that the state shown in Fig.~\ref{Fig:Cluster}a wins if
\begin{equation}
\delta < \frac{C_r}{e^2} \left[(W_{3/4} - W_{1/4})
       - \frac{1}{8} \frac{e^2}{L \kappa}\right].
\label{Condition_on_delta}
\end{equation}
Since $W_{3/4} - W_{1/4} \sim W \sim e^2 / \kappa l_s$ and $L \gg l_s$,
the right-hand side of the inequality~(\ref{Condition_on_delta}) is
positive, and so $\delta > 0$ that satisfy such an inequality do
exist. The only other vaible competing neutral state
is similar to that shown in Fig.~\ref{Fig:Cluster}b except the
`$3/4$-soliton'' is formed on the left semi-infinite segment. The
energy of that state is also $E_b$, and so it does not lead to
any further restrictions on $\delta$.

Now let us examine the charged states. There is only one viable
competitor, with $q = (1 / 2 - \delta) e$, as shown in
Fig.~\ref{Fig:Cluster}c. If the energy difference $\varepsilon = E_c -
E_a$ is positive, then Fig.~\ref{Fig:Cluster}a represents the true
ground state and $\varepsilon$ gives the creation energy of the
charge-$e$ excitation. If $\varepsilon$ is negative, then the ground
state is as shown in Fig.~\ref{Fig:Cluster}c, while the charge and the
energy of the lowest energy charge excitation are equal to $-e$ and
$-\varepsilon$, respectively. An elementary calculation yields
\begin{equation}
\varepsilon = -\frac{e^2}{C_r} \delta + \frac{e^2}{2 \kappa L},
\label{epsilon_cluster}
\end{equation}
from which we conclude that it is possible to obtain arbitrary small
$\varepsilon$ of both signs by tuning $\delta$ sufficiently close to
$C_r / 2 L \kappa \sim c / L$, without violating the
inequality~(\ref{Condition_on_delta}). This proves that clusters are
able to produce a finite $g_B(0)$. In the ground state some of these
clusters are neutral (``empty'') and some are charged (``occupied'').

Now let us try to estimate $g_B$ due to clusters. In the above argument
we required strong inequalities $c \ll l_s$ and $L \gg l_s$ to prove
the existence of a nonzero $g_B$ with mathematical rigor. Physically,
it seems reasonable that such inequalities can be softened to $c \lesssim
l_s$ and $L \gtrsim l_s$, in which case $\delta \lesssim 1$ and $M \sim
1$, i.e., only a few (of the order of unity) impurities are needed. It
is also clear that the rods to the left and to the right of the central
one need not be exactly neutral for the argument to go through. Then the
probability of forming the desired cluster is comparable to the
probability of finding $2 M + 2$ (a few) impurities on the same chain
with nearest-neighbor separation less than $l_s$. Assuming that impurity
positions are totally random and independent, we obtain the estimate of
Eq.~(\ref{g_B_clusters}).

Now let us calculate the form of the Coulomb gap in the actual density
of states $g(\varepsilon)$. At exponentially small energies,
$\varepsilon \ll \varepsilon_* = e^2 / \kappa L_x$, the Coulomb gap is
determined by the interactions on distances exceeding the size of the
Larkin domain. At such distances the interaction has the
form~(\ref{phi_ES}), leading to the usual parabolic Coulomb gap given by
Eqs.~(\ref{CG_ES}) and (\ref{kappa_x}). We put these equations side by
side below for the ease of reading:
\begin{eqnarray}
g(\varepsilon) &=& \frac{3}{\pi} \frac{\kappa^{2}\kappa_x}{e^6}
 \varepsilon^2,\quad
\varepsilon \ll \varepsilon_*,
\label{CG_ES_clean}\\
\varepsilon_* &=& \frac{e^2}{\kappa L_x}
               = \frac{e^2}{\sqrt{\alpha} r_D} \frac{1}{\kappa_x},
\label{epsilon_*}\\
\kappa_x &\sim& \exp\left(C_6 \frac{l}{l_s}\right).
\label{kappa_x_copy}
\end{eqnarray}
To ascertain the region of validity of Eq.~(\ref{CG_ES_clean}) one needs
to make sure that $g(\varepsilon)$ does not exceed the bare density of
states $g_B$. It is easy to see that this is the case here. At the
largest energy $\varepsilon \sim \varepsilon_*$ [Eq.~(\ref{epsilon_*})],
$g(\varepsilon)$ is exponentially small, $g \propto \exp(-C_6 l/l_s)$,
and so it is indeed much smaller than $g_B \sim (l_s / l)^{\beta + 2}$.
In this sense $g_B$ is large enough to ensure the validity of the
parabolic law over the full range of $\varepsilon$ specified in
Eq.~(\ref{CG_ES_clean}).

At $\varepsilon$ larger than $\varepsilon_*$, the Coulomb gap is
governed by interactions within the volume of a Larkin domain and the
dispersion of the dielectric function $\epsilon({\bf q})$ becomes
important. The interaction potential is defined by $\tilde{U}({\bf q}) =
4 \pi e^2 / \epsilon({\bf q}) q^2$ and Eq.~(\ref{epsilon_q_pinned}) in
the $q$-space. In real space, it is given by~\cite{Brazovskii_91}
\begin{eqnarray}
&& U({\bf r}) \simeq \frac{e^2}{2 \kappa |x|} \exp\left(
-\frac{1}{4 \sqrt{\alpha}}
\frac{r_\perp^2}{r_D |x|} \right),
\label{U_paraboloid}\\
&& a_\perp \ll r_\perp \ll {\rm min}\{\sqrt{\alpha} |x|, L_\perp\}, \quad
   \frac{r_D}{\sqrt{\alpha}} \ll x \ll L_x.\quad\quad
\label{U_paraboloid_domain}
\end{eqnarray}
The potential $U({\bf r})$ is appreciable only within the paraboloid
defined by Eq.~(\ref{paraboloid}), where it behaves as $U(r) \simeq e^2 /
2 \kappa |x|$. A more precise statement is that the surface of a
constant $U$ is a paraboloid-like region
\begin{equation}
r_\perp^2 \simeq 4 \sqrt{\alpha} r_D |x|
\ln \left(\frac{e^2}{2 U \kappa |x|}\right).
\label{U_surface}
\end{equation}
For $U \gg \varepsilon_*$, this surface belongs to the
domain~(\ref{U_paraboloid_domain}) where Eq.~(\ref{U_paraboloid}) holds.

To calculate the functional form of the Coulomb gap we use the
self-consistent mean-field approximation due to Efros,~\cite{Efros_76}
according to which $g$ is the solution the integral equation
\begin{equation}
g(\varepsilon) = g_B \exp \left[-\frac12 \int\limits_0^{W}
d \varepsilon^\prime g(\varepsilon^\prime)
V(\varepsilon + \varepsilon^\prime)\right],
\label{Efros_equation}
\end{equation}
where $V(U)$ is the volume enclosed by the constant-$U$
surface~(\ref{U_surface}),
\begin{equation}
V(U) \simeq \frac{\pi \sqrt{\alpha}}{4} \frac{e^4 r_D}{\kappa^2 U^2}.
\label{U_volume}
\end{equation}
Equation~(\ref{Efros_equation}) follows from the requirement that the
ground state must be stable againts a transfer of a unit charge from an
occupied state with energy $-\varepsilon^\prime$ to an empty state with
energy $\varepsilon$. Such a stability criterion~\cite{ES75,ES84} can be
expressed by means of the inequality $\varepsilon + \varepsilon^\prime -
U({\bf r}) > 0$, where ${\bf r}$ is the vector that connects the two
sites. Thus, the integral on the right-hand side of
Eq.~(\ref{Efros_equation}) counts the pairs of states that would violate
the stability criterion if positioned randomly. The Coulomb gap can be
viewed as the reduced statistical weight of stable ground-state charge
distributions with respect to that of the totally uncorrelated ones.

\begin{figure}
\includegraphics[width=1.8in]{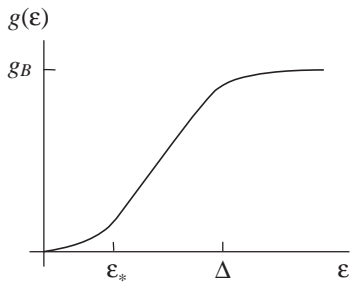}
\vspace{0.1in}
\caption{The density of states of charge excitations in a
3D system with $l \gg l_s$. The parabolic Coulomb gap at low energies is
succeeded by the linear Coulomb gap, then by the bare density of states
$g_B$ created by impurity clusters.
\label{DOS3D}}
\end{figure}

The solution of Eq.~(\ref{Efros_equation}) has the asymptotic
form
\begin{equation}
g(\varepsilon) \simeq \frac{8}{\pi \sqrt{\alpha}}
\frac{\kappa^2}{r_D e^4} |\varepsilon|,\quad
\varepsilon_* \ll |\varepsilon| \ll \Delta.
\label{CG_U} 
\end{equation}
We see that the unusual interaction potential of
Eq.~(\ref{U_paraboloid}) leads to a nonstandard Coulomb gap, which is
linear in a 3D-system. The weaker (linear instead of the standard
quadratic) suppression of $g$ is due to a metallic screening of the
Coulomb potential in the major fraction of solid angle. The energy scale
$\Delta$ in Eq.~(\ref{CG_U}) is defined by the relation $g(\Delta) \sim
g_B$. Using Eq.~(\ref{g_B_clusters}), we can estimate $\Delta$ as
follows:
\begin{equation}
\Delta = \frac{\sqrt{\alpha} r_D}{\kappa^2} e^4 g_B
       \sim W \left(\frac{l_s}{l}\right)^{\beta + 2}.
\label{Delta} 
\end{equation}
At $\varepsilon \gg \Delta$, the solution of Eq.~(\ref{Efros_equation})
approaches the bare density of states, $g \simeq g_B$, and so $\Delta$
has the meaning of the Coulomb gap width. On the lower-energy side, at
$\varepsilon \sim \varepsilon_*$, the linear Coulomb dependence of
Eq.~(\ref{CG_U}) smoothly matches the quadratic dependence of
Eq.~(\ref{CG_ES_clean}). All such dependencies are summarized in
Fig.~\ref{DOS3D}.

%


\subsection{VRH transport in a system with small impurity concentration}
\label{Sec:Transport}
 
The VRH transport at $l \gg l_s$ involves quantum tunneling of charge
excitations between rare impurity clusters. In App.~\ref{Sec:Tunneling}
we advance arguments that the charge tunnels along the chains in the
form of $2\pi$-solitons and derive the following estimate
for the corresponding longitudinal localization length:
\begin{equation}
\xi_x \sim \frac{l_s}{\sqrt{r_s}} \left[1 + \frac{l_s}{l}
\ln\left(\frac{l_s}{a}\right) \right]^{-1},\quad l > l_s.
\label{zeta_clean_exact}
\end{equation}
At $l \gg l_s$ this formula goes over to Eq.~(\ref{zeta_clean}), while
at $l \sim l_s$ it smoothly matches with Eq.~(\ref{zeta_dirty}). As for
the interchain tunneling, it is still accomplished by single-electron
like excitations and the corresponding localization length $\xi_\perp$
is still given by Eq.~(\ref{eta}), see App.~\ref{Sec:Tunneling}. We will
assume that $\xi_\perp / a_\perp$ is not vanishingly small, in which
case the VRH has the 3D character.~\cite{ES_law_1D_hopping}

The density of states and localization lengths is all the information we
need to calculate the VRH conductivity. A more complicated dependence of
the density of states $g(\varepsilon)$ on energy in the case at hand, $l
> l_s$, brings about a larger variety of possible transport regimes.

At the lowest temperatures we still have the ES law with parameter
$T_{\rm ES}$ given by Eq.~(\ref{T_ES_anisotropic}). Due to the exponential
growth of the longitudinal dielectric constant $\kappa_x$ with $l$
[Eq.~(\ref{kappa_x})], $T_{\rm ES}$ decreases exponentially, $T_{\rm ES} \propto
\exp(-C_6 l / 3 l_s)$. This dependence entails a precipitous drop of
$-\ln\sigma$ as a function of $l$, as soon as $l$ exceeds $l_s$, see
Fig.~\ref{sigma3D_l}. Such an enhancement of the conductivity is due to
progressively more efficient screening of the long-range Coulomb
interactions (steeply increasing $\kappa_x$), which enhances the density
of states inside the Coulomb gap, see Eq.~(\ref{CG_ES_clean}).

The ES law~(\ref{DC}) holds until the range of $\varepsilon$'s that
contribute to the VRH transport,~\cite{ES84} $|\varepsilon| \lesssim (T
T_{\rm ES})^{1/2}$, fits inside the parabolic part of the Coulomb gap,
$|\varepsilon| \lesssim \varepsilon_*$. For a fixed $T \ll W$ this gives
the condition $l \lesssim l_s \ln (W / T)$.

At larger $l$, the unusual linear Coulomb gap~(\ref{CG_U}) leads to the
novel 2/5-law for the VRH transport [Eq.~(\ref{DC1})], which we
reproduce below for convenience:
\begin{equation}
\sigma = \sigma_0 \exp[-(T_{1}/T)^{2/5}].
\label{DC1_copy}
\end{equation}
As emphasized in Sec.~\ref{Sec:Results}, parameter $T_1$ is
impurity-independent and is, in this sense, universal, see
Eq.~(\ref{DCT1}). This behavior leads to the intermediate
plateau at the graph in Fig.~\ref{sigma3D_l}.

The range of energies that contributes to the VRH in the 2/5-law regime
is given by $|\varepsilon| \lesssim (T_1 / T)^{2 / 5} T$. At $l \sim l_s
(W / T)^\gamma$, where $\gamma = 3 / [5 (\beta + 2)] \lesssim 0.3$, this
range becomes broader than the Coulomb gap width $\Delta$
[Eq.~(\ref{Delta})]. At such and larger $l$, the Coulomb gap can be
neglected, and the VRH begins to follow the usual Mott law
[Eq.~(\ref{DCM})]
\begin{equation}
\sigma = \sigma_0 \exp[-(T_{M}/T)^{1/4}],
\label{DCM_copy}
\end{equation}
with parameter $T_M$ increasing with $l$ according to $T_M \propto 1 /
g_{B} \propto (l / l_s)^{\beta + 2}$. The growing $T_M$ leads to
exponentially increasing resistivity, represented by the ascending
branch of the curve in Fig.~\ref{sigma3D_l}. Physically, the suppression
of the DC conductivity stems from decreasing density of low-energy
states available for transport, just like in conventional doped
semiconductors~\cite{Mott_book,ES84} or commensurate
CDW systems.~\cite{Kivelson}


\section{Conclusions and comparison with experiment}
\label{Sec:Experiment}

It is widely recognized that interactions must play a significant role
in determining the properties of 1D and quasi-1D conductors, because in
such materials the dimensionless strength of the Coulomb interaction is
very large, $r_s \gg 1$. In the presence of impurities, these systems
behave as insulators and do not possess metallic screening. Thus, the
interactions are both strong and long-range. Our main goal in this paper
was to understand the effect of such interactions on the nature of the
low-energy charge excitations and their Ohmic dc transport. To that end
we formulated a generic model of an anisotropic electron system with
strong Coulomb interactions and disorder and presented its theoretical
analysis. We elucidated the origin of the low-energy charge excitations
in this model and demonstrated that their density of states possesses a
soft Coulomb gap. In 3D case, we found that the Coulomb gap exhibits a
power-law dependence on the energy distance from the Fermi level. We
discussed how the prefactor and the exponents of this power-law
vary as a function of the impurity concentration and other
parameters of the model. We also discussed how the Coulomb gap is
manifested in the variable-range hopping conductivity at low
temperatures.

One of the central results of our theory is a nonmonotonic dependence of
$\sigma$ on the impurity concentration $N$, as shown in
Fig.~\ref{sigma3D_l}, where we sketched $\sigma$ as a function of $l = 1
/ N a_\perp^2$, i.e., as $N$ decreases, at fixed $T$. As clear from that
Figure and the discussion in Sec.~\ref{Sec:Results}, at large $N$ the
conductivity increases as $N$ decreases, similar to behavior found in
metals. In contrast, at small $N$ the conductivity drops as $N$ goes
down, which resembles the behavior of doped semiconductors. For
intermediate $N$, our theory predicts the existence of an
$N$-independent plateau.

\begin{figure}
\includegraphics[width=3.2in]{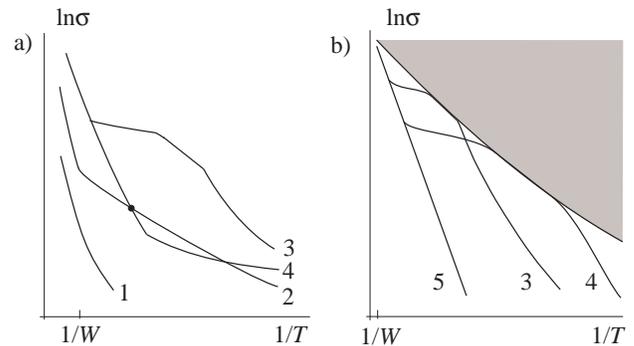}
\caption{Logarithm of the conductivity vs. the inverse temperature for
several samples labelled in the order of increasing $l$, i.e., sample
purity. (a) Curve 1 corresponds to $l = l_1 < l_s$ and displays only the
ES regime. The higher-$T$ activation regime is beyond the limits of the
graph. Curve 2 is for $l = l_2 \gtrsim l_s$, and so the both the
activation behavior and the ES law are visible. Curve 3 is for $l = l_3
\gtrsim l_2$. It shows the complete sequence of the transport regimes:
the activation, Mott, $2/5$, and ES laws. Curve 4 depicts the behavior
of a sample with an impurity concentration another notch lower than that
of sample 3. In panel (a) it goes through the activation and the Mott
regimes. In panel (b) that covers considerably lower $T$, curve 4 also
exhibits the $2/5$ law, followed by the ES law. The $2/5$-law is shown
by the solid line along the lower edge of the shaded region. The curves can
skim along this line but cannot cross it. The higher the sample purity
the lower the temperature at which the sample starts to exhibit the
$2/5$-law but also the wider the range of $T$ over which this law
persists. Curve 5 depicts the case $l \gg l_s$ where any kind
of VRH transport corresponds to Ohmic resistances higher than the
experimental measurement limit, so that only the activated transport can be
observed.
\label{sigma_cross}}
\end{figure}

Another way to represent these theoretical predictions, common in
semiconductor physics, is shown in Fig.~\ref{sigma_cross}. In that
Figure the dependence of the logarithm of conductivity on the inverse
temperature is depicted for a series of samples, each with fixed $N$. An
unusual circumstance illustrated by Fig.~\ref{sigma_cross}, is the
crossing of the curves that correspond to different samples. In
Fig.~\ref{sigma_cross}a we show that up to two crossing points may exist
between one curve that corresponds to $l < l_s$ (curve 2) and another
curve for $l > l_s$ (curve 4). The higher-$T$ crossing occurs when the
curve 4 goes through the activation regime, the lower-$T$ one --- when
it exhibits the Mott VRH. The dirtier ($l < l_s$) sample obeys the ES
behavior in both instances. Another property we tried to emphasize in
Fig.~\ref{sigma_cross} is the role of the 2/5-law as the upper bound of
the conductivity regardless of the the sample purity. For samples with
low impurity concentration the 2/5-law is also the envelope curve, see
Fig.~\ref{sigma_cross}b.

Let us now turn to the experimental situation. The transport behavior of
a number of organic compounds, including TMTSF-DMTCNQ, TTF-TCNQ, and
NMe-4-MePy (TCNQ)$_2$ is indeed in a qualitative agreement with our
theory. It should be clarified that such materials form CDW phases that
in addition to the usual $2 k_F$ periodicity, have an appreciable or, in
some cases, even predominant $4 k_F$ harmonics. The latter is considered
the evidence for the strong Coulomb interaction,~\cite{Organics_review}
and so is precisely the case we studied in this paper.

In the experiments of Zuppiroli {\it et
al.\/}~\cite{Zuppiroli_80,Zuppiroli_review} the transport in the
aforementioned compounds was studied as a function of defect
concentration, which was varied {\it in situ\/} by irradiation of
samples by high-energy particles. Admittedly, the nature of the such
defects is not known with certainty. Some suggestions in the literature
include atomic displacements, broken bonds, polymer cross-linking, and
charged radicals. At the same time, the effect of the irradiation on
transport seem not to depend much on the type of particles used (X-rays,
neutrons, or electrons) and instead to correlate primarily with the
total absorbed energy.~\cite{Zuppiroli_review} This fact is interpreted
as evidence that microscopically defferent defects influence the
transport in electron crystals in a similar way, so long as they act as
strong localized pinning centers. Under this assumption, it is
legitimate to compare the data from the irradiation experiments with our
theory even though so far we assumed that defects are created by charged
acceptors (see Sec.~\ref{Sec:Introduction}). We do so in some detail
below.

\begin{figure}
\includegraphics[width=2.7in,bb=71 251 531 629]{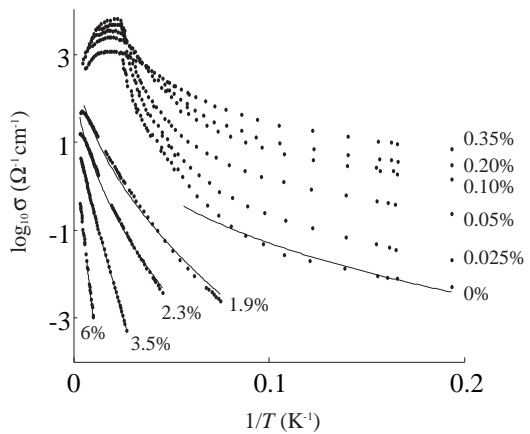}
\caption{Low-temperature conductivity of TMTSF-DMTCNQ samples damaged by
X-ray radiation. The dots on $c \geq 1.9\%$ and $c \leq 0.35\%$ curves were
generated by digitizing the experimental data in Fig.~1 of
Ref.~\onlinecite{Zuppiroli_JPhys_80} and Fig.~2 of
Ref.~\onlinecite{Zuppiroli_review}, respectively. The percentages stand
for the defect concentrations quoted in those papers. The solid line through
the $6\%$ and $3.5\%$ data are the best fits to the activation; through
the $2.3\%$ and $1.9\%$ --- to the ES law. In both cases the prefactors
($\sigma_A$ and $\sigma_0$) were taken to be $T$-independent. The thin
line through the $0\%$ data (unirradiated sample) is the best fit to the
Mott law based on the $T < 20\,{\rm K}$ points.
\label{Fig:Zuppiroli_fits}}
\end{figure}

In Fig.~\ref{Fig:Zuppiroli_fits} we show an extensive set of data on
transport in irradiated TMTSF-DMTCNQ that we assembled by digitizing
Fig.~2 of a review article by Zuppiroli~\cite{Zuppiroli_review} and
original references therein. Apparently, some data series in this figure
represent the same sample with successively increasing radiation dose,
and some correspond to physically different specimens. For simplicity,
we refer to all of them as different samples. The percentage labels on
the plot are the estimates of the molar concentrations of defects ($c$)
quoted by the experimentalists. The points on the $c = 0$ trace are
from an unirradiated sample.

As shown in Fig.~\ref{Fig:Zuppiroli_fits}, the data for the two most
disordered samples can be successfully fitted to the activation law and
the next two samples --- to the ES formula. This transition from the
activation to the ES law with increasing disorder is in agreement with
our theory (cf. curves 1 and 2 in Fig.~\ref{sigma_cross}). The obtained
fit parameters $E_A$ and $T_{\rm ES}$ are given in
Table~\ref{Tab:Zuppiroli_fits}. Both $E_A$ and $T_{\rm ES}$ scale roughly
linear with $c$, in agreement with Eqs.~(\ref{E_A}) and~(\ref{DCT}).
From Eq.~(\ref{DCROD}) we deduce that $\kappa l c = \kappa a \sim
1\,{\rm nm}$, which has the correct order of magnitude (assuming $\kappa
\sim 1$). One should keep in mind here that the absolute numbers for $c$
were obtained by the authors of Ref.~\onlinecite{Zuppiroli_JPhys_80}
using certain arguable assumptions. In our opinion, the scaling with $c$
may be more reliable than the absolute values quoted because (if no
annealing occurs) the relative magnitude of $c$ should scale linearly
with the irradiation time, known to experimentalists without any fitting
parameters. Combining Eqs.~(\ref{DCT}) and~(\ref{E_A}) we further
deduce that $a_\perp /\xi_\perp$ and $l / \xi_x$ ratios are some modest numbers
less than ten, as may be expected from Eqs.~(\ref{zeta_dirty})
and~(\ref{eta}). 

As a final remark on high-disorder samples, we would like to mention
that the scaling of the longitudinal dilectric constant $\kappa_x$ with
the defect concentration (irradiation time) consistent with
Eq.~(\ref{kappa_x_dirty}) was reported in a separate set of experiments
on Qn(TCNQ) by Janossy {\it et al.\/}~\cite{Janossy_82} Together with
the transport data, this makes a compelling case for the
validity of the metallic-rod (interrupted-strand) model for organic
electron crystals with $l \ll l_s$. For such systems we can claim
a semi-quantitative agreement with the experiment.

\begin{table}
\caption{\label{Tab:Zuppiroli_fits} Fitting parameters for data displayed
in Fig.~\ref{Fig:Zuppiroli_fits}.}
\begin{ruledtabular}
\begin{tabular}{cccc}
Quoted defect concentration, $c$ & $T_{\rm ES}$, K  & $E_A$, K & \\ 
\hline
$\sim 6 \%$                      &   --         & 900\\
 $3.5 \%$                        &   --         & 380\\
 $2.3 \%$                        &  2800        &  -- \\
 $1.9 \%$                        &  3700        &  --        
\end{tabular}
\end{ruledtabular}
\end{table}

Let us now turn to the conductivity of weakly damaged
samples.~\cite{Zuppiroli_review} As one can see from
Fig.~\ref{Fig:Zuppiroli_fits} they show metallic behavior at high-$T$, a
conductivity maximum at the Peierls temperature of about $42\,{\rm K}$,
and a decrease in $\sigma$, i.e., the semiconducting behavior, at lower
$T$. As $T$ drops by a factor of two or so below the Peierls
temperature, the decrease of $\sigma$ with $1 / T$ becomes considerably
more gentle than an activation law. One can not help noticing a
similarity between the behavior of $c = 1.9, 0.35$, and $0\%$ samples in
Fig.~\ref{Fig:Zuppiroli_fits} and curves 1, 3, and 4, respectively, in
Fig.~\ref{sigma_cross}. There is also an unambiguous evidence for the
existence of the crossing point between, e.g., $c = 0\%$ and $c \approx
1.2\%$ traces at $T = 21\,{\rm K}$ (see below). However, an attempt to
fit the $c = 0\%$ data to the Mott law is not particularly successful,
see Fig.~\ref{Fig:Zuppiroli_fits}. Therefore, we only wish to emphasize
a qualitative agreement with our prediction that for a fixed $T$, the
conductivity of ``clean'' and ``dirty'' samples should show opposite
trends, see Figs.~\ref{sigma3D_l} and~\ref{sigma_cross}. Indeed, the
conductivity of the low-disorder samples ($c \leq 0.35\%$) increases
with the radiation dose in contrast to the behavior shown by the
high-disorder samples ($c \geq 1.9\%$) where it decreases. In fact,
another experiment showed this contrasting behavior in a great detail.
In that experiment,~\cite{Zuppiroli_JPhys_80} $\ln\sigma$ was measured
at the fixed temperature of $T = 21\,{\rm K}$, while defect
concentration was varied essentially continuously over the range of $0 <
c < 2.5\%$. It was found that $\sigma$ initially increases by two orders
of magnitude, reaches a maximum, and then drops by five orders of
magnitude as a function of $c$. Overall, this is in a qualitative
agreement with Fig.~\ref{sigma3D_l} except instead of the well-defined
$2/5$-plateau, $\ln\sigma$ shows only a broad maximum. Similar features
are demonstrated also by TTF-TCNQ and NMe-4-MePy (TCNQ)$_2$, see Fig.~1
in Ref.~\onlinecite{Zuppiroli_80}, and Figs.~11 and 12 in
Ref.~\onlinecite{Zuppiroli_review}. We conclude that for low-disorder
CDW organics our theory agrees with experiment in some gross
qualitative features. The quantitative agreement cannot be verified
because the dynamical range of measured conductivities is too narrow.
Further low-temperature experiments are desired to clarify the
situation and to prove or disprove the existence of the novel $2/5$-law.
 
Let us now switch to inorganic CDW. Several comments are in order. The
electron-electron interactions in these materials are also very
strong,~\cite{Gruner,Artemenko_95} $r_s \sim 100$. However, inorganic
CDW are predominantly $2 k_F$, and there is an ample evidence for the
important role of the electron-phonon coupling in the CDW dynamics. This
coupling can lead to an enhancement of the electron effective
mass,~\cite{Lee_74,Gruner} that would result in a short localization
length. If the mass enhancement is indeed large, the VRH transport
should be observable only in materials with short hopping distances,
i.e., large impurity concentrations. Examples include highly doped
bronzes~\cite{Rouziere_00} and perhaps, the ${\rm Pt}$-chain compound
KCP.~\cite{Zeller_71,Thomas} From this perspective, the reports of a
VRH-like transport in a relatively pure samples of ${\rm Ta}{\rm S}_3$
[Refs.~\onlinecite{Zhilinskii_83} and~\onlinecite{Itkis_90}] and blue
bronze [Refs.~\onlinecite{Hosseini_99} and~\onlinecite{Zawilski_02}] are
puzzling and require further investigations.

Finally, let us comment on another broad class of quasi-1D systems,
conducting polymers. In comparison to CDW, polymers have a much higher
degree of structural disorder and a complex morphology that depends on
the preparation method. Typical samples contain of a mixture of
crystalline and amorphous regions, with the correlation
length~\cite{Joo_98} of the order of $10\,{\rm nm}$. In the undoped
state polymers are commensurate CDW semiconductors with a Peierls-Mott
energy gap~\cite{Heeger_84} $\sim 1\,{e V}$. Doping shifts these systems
away from commensurability point and suppresses the gap but it is often
inhomogeneous and is an additional source of disorder. At low and
moderate doping the $T$-dependence of the conductivity often resembles
ES and/or Mott VRH laws, see a short review in
Ref.~\onlinecite{Yoon_95}. A systematic study of the VRH conductivity
dependence on doping has been attempted by Aleshin~{\it et
al.\/}~\cite{Aleshin_97} In those experiments doping of PEDT/PSS samples
was varied by controlling the pH of the solution at the sample
preparation stage. It was observed that at ${\rm pH} < 4$ the VRH
exponent $\lambda$ [Eq.~(\ref{VRH_exponent})] was close to 0.5 and
$T_{\rm VRH}$ decreased with pH, while at larger ${\rm pH}$, $\lambda$
was close to 0.4 = 2/5 and $T_{\rm VRH}$ did not depend on pH. This
resembles the behavior that follows from our theory, provided the
concentration of the pinning centers decreases with pH. We leave the
tasks of extending our theory to the case of conducting polymers and
explaning these intriguing experimental results for future
investigations.

\acknowledgments

This work is supported by Hellman Scholarship Award at UCSD (M.~M.~F.),
NSF Grant No.~DMR-9985785 (B.~I.~S.), and INTAS Grant No.~2212 (S.~T.).
We thank S.~Brazovskii, A.~Larkin, and S.~Matveenko for valuable
comments, K.~Biljakovic, O.~Chauvet, S.~Ravy, D.~Staresinic, and
R.~Thorne for discussions of the experiments, and Aspen Center for
Physics, where a part of this work was conducted, for hospitality.

\appendix

\section{Dielectric function and elasticity in a clean quasi-1D electron crystal}
\label{Sec:Elasticity}

In this section we derive expressions for the elastic moduli and the
dielectric function of a pure crystal.

Following the literature on interacting 1D electron
systems~\cite{Gogolin_book} and CDW,~\cite{Lee_74} we describe
dynamics of electrons on $i$th chain by a bosonic phase field
$\varphi_i(x, t)$. The the long-wavelength components of electron
density $n_i$ is related to $\varphi_i$ by $n_i = \partial_x \varphi_i /
2 \pi$. The elasticity theory of the system can be formulated by
identifying the elastic displacement $u$ with $(2 \pi / a) \varphi$ and
taking the continuum limit. Neglecting weak interchain tunneling and
dynamical effects we choose our starting effective Hamiltonian $H$ in
the form
\begin{eqnarray}
&H = \int d x [H_{0} + H_{C}],&
\label{H}\\
&\displaystyle H_{0} = \frac{1}{2} C_x^0
 \sum\limits_{i} n_i^2
+
 \sum\limits_{i j} J_{i j} \cos(\varphi_i - \varphi_j),&
\label{H_0}\\
&H_{C} =  \frac12 \sum\limits_{i j} \int d x d x^\prime
n_i(x) U_{i j}(x - x^\prime) n_j(x^\prime),&
\label{H_C}
\end{eqnarray}
Let us briefly describe the notations here. The Hamiltonian is split
into the short-range ($H_0$) and the long-range Coulomb ($H_C)$ parts.
$J_{i j}$ represents the Coulomb coupling between the CDW modulations of
electron density on chains $i$ and $j$. $C_x^0$ is the charge
compressibility of a single isolated chain. In a large-$r_s$ 1D system
$C_x^0$ is dominated by the exchange-correlation effects,
leading to~\cite{Fogler_xxx}
\begin{equation}
C_x^0 \sim -\frac{2 e^2}{\kappa} \ln \frac{a}{R}
\label{C_x_0}
\end{equation}
where $R$ is the characteristic radius of the electron charge
form-factor in the transverse direction, i.e., the ``radius'' of the
chain. In most physical realizations, we expect $R \ll a$ and {\it
negative\/} $C_x^0$. The positivity of the elastic modulus of the
system, required for thermodynamic stability, is recovered once we take
into account the long-range part $H_C$ of the interaction energy,
parametrized by the kernels $U_{i j}$. $U_{i j}$ is defined as the bare
Coulomb kernel $U_0(r) = e^2 / \kappa r$ convoluted with the
single-chain form-factors $F$, e.g., $F({\bf q}_\perp) =
\exp(-q_\perp^2 R^2)$.

We are interested in a linear response where the cosines in
Eq.~(\ref{H_0}) can be expanded in $\varphi$ thereupon the effective
Hamiltonian becomes quadratic. If an external electrostatic potential
$\tilde{V}_{\rm ext}({\bf q})$ acts on the system, the total equilibrium
potential $\tilde{V}({\bf q})$ will in general contain Fourier harmonics
with wavevectors ${\bf q} + {\bf G}$ where ${\bf G}$ are the reciprocal
vectors of the 2D lattice formed by the transverse coordinates of the
centers of the chains. We define the dielectric function $\epsilon({\bf
q})$ of the system as the ratio $\tilde{V}_{\rm ext}({\bf q}) /
\tilde{V}({\bf q})$ for ${\bf q}$ in the Brillouin zone of this lattice.
Via standard algebraic manipulations in the reciprocal space we arrive
at
\begin{eqnarray}
\epsilon({\bf q}) &=& \kappa + \frac{4 \pi e^2}{a_\perp^2}
\frac{q_x^2}{q^2} \frac{1}{B_x({\bf q}) q_x^2 + B_\perp q_\perp^2},
\label{epsilon_q_I}\\
B_x({\bf q}) &=& C_x^0 + \frac{4 \pi e^2}{\kappa a_\perp^2}
 \sum_{{\bf G} \neq 0} \frac{F^2({\bf G})}
     {q_x^2 + ({\bf q}_\perp + {\bf G})^2},
\label{B_x}\\
B_\perp &=& 4 \pi^2 a_\perp^2 \sum_j J_{i j}.
\label{B_perp}
\end{eqnarray}
In the limit $q_x \ll a^{-1}$, $q_\perp \ll a_\perp^{-1}$ we obtain
Eq.~(\ref{epsilon_q}) reproduced below for convenience,
\begin{equation}
\epsilon({\bf q}) = \kappa + \frac{q_x^2}{q^2}
\frac{\kappa r_D^{-2}}{q_x^2 + \alpha q_\perp^2}.
\label{epsilon_q_copy}
\end{equation}
Here $r_D = (C_x \kappa / 4 \pi e^2)^{1/2} a_\perp$ is the Thomas-Fermi screening
radius, $C_x > 0$ is the effective charge compressibility of the system,
\begin{equation}
C_x = C_x^0 + \frac{4 \pi e^2}{\kappa a_\perp^2} \sum_{{\bf G} \neq 0}
\frac{F^2({\bf G})}{G^2}
\sim \frac{2 e^2}{\kappa} \ln \frac{a_\perp}{a},
\label{C_x}
\end{equation}
and $\alpha = B_\perp / C_x$ is the dimensionless anisotropy parameter.
Let us now discuss some consequences of Eq.~(\ref{epsilon_q_copy}).

(i) The dielectric function has the same form as in quasi-1D
systems with small $r_s$ (see, e.g., Ref.~\onlinecite{Brazovskii_93}).

(ii) The screening radius $r_D$ is of the order of the interchain
separation $a_\perp$. Therefore, Brazovskii and Matveeenko's
results~\cite{Brazovskii_91} for the soliton energy $W$ and the soliton
length $l_s$ remain qualitatively correct~\cite{Comment_on_Artemenko} for
high $r_s$ provided we use $r_D \sim a_\perp$.

(iii) The interaction energy $U({\bf r})$ of two point-like test charges
separated by a large distance ${\bf r}$ can be calculated by the Fourier
inversion of
\begin{equation}
\tilde{U}({\bf q}) = \frac{4 \pi e^2}{\epsilon({\bf q}) q^2}.
\label{U_q}
\end{equation}
For $\alpha x^2 + r_\perp^2 \gg r_D^2$ one finds
(cf.~Ref.~\onlinecite{Brazovskii_91})
\begin{equation}
U({\bf r}) \simeq \frac{e^2}{2 \kappa |x|} \exp\left(
-\frac{1}{2 r_D}
\frac{r_\perp^2}{\sqrt{\alpha x^2 + r_\perp^2} + \sqrt{\alpha} |x|}
\right).
\label{U}
\end{equation}
The potential $U({\bf r})$ is not exponentially small only within the
paraboloid $r_\perp^2 \lesssim \sqrt{\alpha} r_D |x|$
[cf.~Eq.~(\ref{paraboloid})]. At the surface of such a paraboloid we
have $r_\perp \ll \sqrt{\alpha} |x|$ and $U({\bf r})$ acquires a simpler
form quoted in Sec.~\ref{Sec:3D_clean},
\begin{equation}
U({\bf r}) \simeq \frac{e^2}{2 \kappa |x|} \exp\left(
-\frac{1}{4 \sqrt{\alpha}}
\frac{r_\perp^2}{r_D |x|} \right).
\label{U_paraboloid_copy}
\end{equation}

(iv) Finally, the effective longitudinal and transverse
elastic moduli of the system are given by
\begin{eqnarray}
&\displaystyle Y_x = \frac{C_x}{a^2 a_\perp^2}
 \sim \frac{e^2}{\kappa a^2 a_\perp^2},&
\label{Y_x_copy}\\
&\displaystyle Y_\perp = \alpha Y_x.&
\label{Y_perp}
\end{eqnarray}
%


\section{Tunneling in a 3D crystal with a low impurity concentration}
\label{Sec:Tunneling}

The localization lengths $\xi_\perp$ and $\xi_x$ needed for calculation
of the VRH transport are determined by long-distance tunneling of charge
excitations. The problem of tunneling is nontrivial because a broad
spectrum of charge excitations exists. Leaving more detailed
investigations for future work, we concentrate on two possible tunneling
mechanisms: by electron-like quasiparticles and by many-body
excitations, the $2\pi$-solitions.

In the quasiparticle mechanism the charge is carried by a single
electron while all other electrons remain unperturbed in their quantum
ground states. The rational for examining this mechanism is its minimal
possible tunneling mass. The problem of calculating $\xi_x$ and
$\xi_\perp$ reduces to the quantum mechanics of a single particle in a
fixed external potential. Clearly, the optimal tunneling path should go
through the interstitial positions where the energy barrier is the
lowest. It is convenient then to formulate the problem as a problem on a
lattice of such interstitial positions. The relevant variables are the
on-site energies and the hopping matrix elements. The on-site energies
are all equal to $\varepsilon_{\rm int} \sim e^2 / \kappa a$. The
hopping terms for the interchain tunneling, $t_\perp$, are determined by
the band-structure in the case of tunneling inside a chemically
synthesized materials. In the case of tunneling between distant 1D
conductors $t_\perp \propto \exp(-a_\perp / a_B)$. The hopping matrix
element for tunneling along the chain, $t_x$, can be estimated
straightforwardly, with the result $t_x = \varepsilon_{\rm int}
\exp(-C_7 \sqrt{r_s})$, $C_7 \sim 1$. In the absence of impurities, the problem is
reduced to the propagation through a periodic lattice. The eigenstates
in that model are labelled by wavevectors ${\bf k}$, according to the
tight-binding dispersion relation
\[
\varepsilon_{tb}({\bf k}) = \varepsilon_{\rm int} - 2 t_x \cos (k_x a_\perp)
 - 2 t_\perp [\cos (k_y a_\perp) + \cos (k_z a_\perp)].
\]
Below the band edge $\varepsilon = \varepsilon_{\rm int} - 2 t_x - 4
t_\perp$, the eigenstates are exponentially decaying with distance. The
corresponding localization (decay) lengths can be related to the
imaginary parts of the complex ${\bf k}$ solution of the equation
$\varepsilon_{tb}({\bf k}) = \varepsilon$. In the case of interest,
$\varepsilon \ll \varepsilon_{\rm int}$; $t_\perp, t_x \ll
\varepsilon_{\rm int}$, we obtain
\begin{eqnarray}
\xi_x^q &=& \frac{a}{\ln (\varepsilon_{\rm int} / 2 t_x)}
\sim \frac{a}{\sqrt{r_s}},
\label{xi_x_q}\\
\xi_\perp^q &=& \frac{a_\perp}{\ln (\varepsilon_{\rm int} / 2 t_\perp)},
\label{xi_perp_q}
\end{eqnarray}
where the superscript $q$ stands for ``quasiparticle.'' Although
Eq.~(\ref{xi_perp_q}) was derived for a clean system, it is clear that
impurity do not affect this result unless present in gigantic
concentrations ($l \sim a$). Indeed, an individual impurity can modify
the local on-site energy by at most a numerical factor, while
$\xi_{\perp (x)}$ depend on the on-site energy only weakly,
logarithmically. In principle, impurity clusters with atypically low
on-site energies, resonant with $\varepsilon$, do exist but as well
known from the analysis of the resonant tunneling problem in random
systems, such events are exponentially rare and do not contribute to the
bulk localization length in any appreciable manner.

Let us now turn to the solition mechanism, we we attempt to profit from
the fact that in the bulk, the soliton is the charge excitation of the
lowest possible energy, so that the energy barrier could perhaps be
lower and the tunneling more effective. However, in the case of
interchain tunneling, this is not the case. Indeed, the direct tunneling
of a soliton to a different chain is impossible because the soliton is a
composite many-body excitation. The closest to the interchain soliton
tunneling that one can imagine is a two-stage process, where, one
electron first tunnels to the adjacent chain, and then, on the second
stage, it pushes away other electrons in the region of length $l_s$ to
form a soliton. Since the initial energy barrier is still
$\varepsilon_{\rm int}$ and the charge spreading only increases the
tunneling action, it is clear that such a contrived process offers no
advantage compared to the simple one-stage quasiparticle mechanism.
Therefore, $\xi_\perp$ is determined by the latter and coincides with
$\xi_\perp^q$, leading to Eq.~(\ref{eta}). Note that for the case of
{\it distant\/} chains where $t_\perp \sim \exp(-a_\perp / a_B)$, the correct
limiting result $\xi_\perp = a_B$ is recovered.

In contrast, for the tunneling along the chain the soliton mechanism is
the winner. Consider first the longitudinal tunneling of a
soliton in the absence of impurities. Empoying the usual
imaginary-time picture, the action for tunneling over a distance $x \gg
l_s$ can be estimated as the product of the energy barrier $W$ and the
tunneling time $\tau \sim x / u$. Here $u$, given by
\begin{equation}
u = \left(\frac{C_x}{m a}\right)^{1/2} \sim \frac{e^2}{\kappa \hbar}
\frac{1}{\sqrt{r_s}},
\label{u}
\end{equation}
is the sound velocity.~\cite{Comment_on_u} 
The tunneling amplitude is of the order of $\exp(-W x / \hbar u)$.
Using $W \sim e^2 / \kappa l_s$, we arrive
at the estimate of the localization length as follows
\begin{equation}
\xi_x^s \sim \frac{l_s}{\sqrt{r_s}}\quad (l = \infty),
\label{xi_x_s}
\end{equation}
where the superscript $s$ stands for ``soliton.'' Clearly, $\xi_x^s \gg
\xi_x^q$, so that the soliton mechanism dominates the longitudinal
tunneling. To account for the dilute impurities, we should add to the
above expression for the action an extra term $(x / l) [\hbar \sqrt{r_s}
\ln(l_s / a)]$. Here the factor $(x / l)$ is the average number of
impurities on the tunneling path of length $x$ and the expression inside
the square brackets is the action cost for compactification of the
charge-$e$ from the length $l_s$ to length $a$ and spreading it back
during the tunneling through each impurity. In this manner, we obtain a
corrected expression for $\xi_x$, which coincides with
Eq.~(\ref{zeta_clean_exact}).


\section{Dimensional energy estimates for the collective pinning}
\label{Sec:Collective}

In this Appendix we use the ideas of collective pinning to derive the
growth of elastic distortions in a quasi-1D crystal pinned by strong
dilute impurities. We also derive the estimates of the corresponding
gain in the pinning energy density.

We start by reformulating the argument leading to Eq.~(\ref{larkin_l})
in the language conventional in the literature devoted to collective
pinning.~\cite{Blatter_94} To do so we note that since the energy of a
given soliton dipole $E_s$ depends on the background elastic
displacement field $\bar{u}$, each impurity exerts a force $f =
-\partial E_s / \partial \bar{u} \sim W / a$ on the crystal. The
long-range variations of $\bar{u}$ appear in response to such random
forces. Let $\Delta u(D)$ be a characteristic variation of $\bar{u}$
over a distance $D$ in the transverse direction and let $X$ be a typical
distance over which a variation of the same order in the $x$-direction
builds up. Our next step is to estimate the total energy $E$ of a volume
$V = X \times D \times D$ (relative to the energy of a pristine
crystal).

The energy consists of elastic, Coulomb, and pinning parts,
\begin{equation}
                 E = E^{\rm el} + E^C + E^{\rm pin}.
\label{E_split}
\end{equation}
In its turn, $E^{\rm el}$ is the sum of the longitudinal
and transverse terms,
\begin{equation}
E^{\rm el} \sim Y_x \left(\frac{\Delta u}{X}\right)^2 V
 + Y_\perp \left(\frac{\Delta u}{D}\right)^2 V.
\label{E_el}
\end{equation}
The Coulomb energy is of the order of $\tilde{U}(q_x, q_\perp) \rho^2
V$, where $\tilde{U}(q) = 4 \pi e^2 / \kappa q^2$ is the Coulomb kernel,
$q_x = 1 / X$, $q_\perp = 1 / D$ are the characteristic wavevectors
involved, $\rho = e n \partial_x u \sim e n \Delta u / X$ is the
charge density associated with the londitudinal compression, and $n = 1
/ a a_\perp^2$ is the average electron concentration. Below we show
that $D \ll X$, so that $q_\perp \gg q_x$,
$\tilde{U}(q) \sim e^2 D^2 / \kappa$, and finally,
\begin{equation}
   E^{C} \sim \frac{e^2}{\kappa} \frac{D^4}{X}
              \left(\frac{\Delta u}{a a_\perp^2}\right)^2.
\label{E_C}
\end{equation}
The pinning energy can be estimated as $E^{\rm pin} \sim -\Delta u \sum_j
f_j$, where $f_j \sim W / a$ is the force exerted on the lattice by
$j$th impurity. The average number of impurities in the volume $V$ is
$N_i = N V$ and $f_j$ have random signs; hence, $E^{\rm pin} \sim -(W /
a) \Delta u \sqrt{N_i}$, or
\begin{equation}
   E^{\rm pin} \sim -W \left(\frac{\Delta u}{a}\right)
              \left(\frac{D}{a_\perp}\right) \left(\frac{X}{l}\right)^{1/2}.
\label{E_pin}
\end{equation}
Combining Eqs.~(\ref{Y_x_copy}) and (\ref{E_el}-\ref{E_pin}), we arrive
at
\begin{eqnarray}
E &=& \frac{W}{a} \left(\frac{\Delta u}{a_\perp}\right)^2
\left(\alpha X + \frac{D^4}{X a_\perp^2}\right)
\nonumber\\
&-& W \left(\frac{\Delta u}{a}\right) \left(\frac{D}{a_\perp}\right)
\left(\frac{X}{l}\right)^{1/2}.
\label{E}
\end{eqnarray}
$X$ and $\Delta u$ can now be found by optimizing $E$ for a fixed $D$.
Not surprisingly, we find that $X$ and $D$ are related by the defining
equation~(\ref{paraboloid}) of the paraboloid introduced in
Sec.~\ref{Sec:Pinning} (for $r_D \sim a_\perp$ case),
\begin{equation}
X \sim \frac{D^2}{a_\perp \sqrt{\alpha}}.
\label{X}
\end{equation}
Such a paraboloid is an invariable feature of the elastic response of
the quasi-1D crystal to external forces. What is surprising however is
that $\Delta u^2 \sim ({l_s} / {l}) a^2$ is small and does not depend on
$D$, at odds with Eq.~(\ref{u_squared}). The resolution of this
contradiction comes from a realization that what $\Delta u(D)$ really
represents is the elastic distortion due to the adjustment of the
crystal on a single scale $D$. In fact, there is a hierarchy of smaller
scales $D, D / 2, D / 4, \ldots, r_\perp^{\rm min}$, on which adjustments
are approximately independent. The correct estimate of $\Delta u$,
Eq.~(\ref{u_squared}), is obtained once we sum over all such scales,
$\Delta u^2 \sim M a^2 l_s / l$, where $M \sim \ln (D / r_\perp^{\rm min})$
is the number of scales. Thereby, we recover Eq.~(\ref{u_squared}) and
as an additional benefit, we find the expression for the energy of the
collective pinning,
\begin{equation}
E \sim -W \left(\frac{\Delta u}{a}\right)^2 \left(\frac{D}{a_\perp}\right)^2.
\label{E_I}
\end{equation}
Let us define the pinning energy density by ${\cal E}_{\rm pin} = E /
X D^2$. Using Eqs.~(\ref{Y_x}), (\ref{larkin_l}), (\ref{X}), and
(\ref{E_I}) we obtain the estimate of ${\cal E}_{\rm pin}$ at the
Larkin scale as given by Eq.~(\ref{epsilon_pin}). As clear from this
derivation, both Eq.~(\ref{u_squared}) and (\ref{epsilon_pin}) are
essentially the lowest-order perturbation theory results. It is
generally expected that the growth of $\Delta u^2$ with $r$ slows down
beyond the Larkin length~\cite{Comment_on_beyond_Larkin} and that
adjustments on larger scales do not lead to any substantial increase in
the pinning energy density. In this case, Eq.~(\ref{epsilon_pin}) is the
final estimate of ${\cal E}_{\rm pin}$ in the thermodynamics limit.



\begin{thebibliography}{99}
\bibitem{Comment_on_Artemenko} Our conclusions differ from those of
Ref.~\onlinecite{Artemenko_95}, where an unphysical limit $r_D
\ll a_\perp$ is considered.
\bibitem{Comment_on_u} If the electron-phonon coupling is important,
Borh radius $a_B$ and the $r_s$ parameter must be calculated based on
the enhanced electron effective mass~\cite{Lee_74,Gruner}
$m_*$, i.e., $r_s = e^2 a m_* / 2 \hbar^2 \kappa$.
Thus, the mass enhancement decreases the sound velocity $u$ and
supresses the tunneling, as may be expected on physical grounds.
\end{thebibliography}
\end{document}